%Paper: hep-th/9306002
%From: ghoshal@physics.rutgers.edu (Subir Ghoshal)
%Date: Tue, 1 Jun 93 22:05:43 EDT

\input psfig
\magnification 1200
\rightline{RU-93-20}
\vskip 0.4in
\centerline{\bf BOUNDARY S-MATRIX AND BOUNDARY STATE}
\centerline{\bf IN TWO-DIMENSIONAL INTEGRABLE QUANTUM FIELD THEORY}
\vskip 0.6in
\centerline{Subir Ghoshal\footnote{$^\dagger$}{E-mail:
GHOSHAL@ruhets.rutgers.edu} and
Alexander Zamolodchikov \footnote{$^\ddagger$}{E-mail:
SASHAZ@ruhets.rutgers.edu}
\footnote{$^\star$}{Supported in part by the
Department of Energy under grant No. DE--FG0590ER40559}}
\vskip 0.1in
\centerline{Department of Physics and Astronomy}
\centerline{Rutgers University}
\centerline{P.O.Box 849, Piscataway, NJ 08855-0849}
\vskip 0.4in
\centerline{\bf Abstract}

We study integrals of motion and factorizable S-matrices in
two-dimensional integrable field theory with boundary. We propose the
``boundary cross-unitarity equation'' which is the boundary analog of the
cross-symmetry condition of the ``bulk'' S-matrix. We derive the boundary
S-matrices for the Ising field theory with boundary magnetic field and
for the boundary sine-Gordon model.

\vskip 0.4in

1.INTRODUCTION

In this paper we study two-dimensional integrable field theory with
boundary. Exact solution to such a field theory could provide better
understanding of boundary-related phenomena in statistical systems
near criticality[1]. Quantum field theory with boundary can be applied to
study quantum systems with dissipative forces[2]. From a more general
point of view, studying integrable models could throw some light on the
structure of the ``space of boundary interactions'', the object of
primary significance in open string field theory[3].

An integrable field theory posesses an infinite set of mutually commutative
integrals
of motion\footnote{$^1$}{With some reservations, this property is widely
believed to be a sufficient condition. In practice, it is usually enough
to find the first few integrals of motion of spin $s>1$ to argue about
integrability (this argument works in all known cases).}. In the ``bulk
theory'' (i.e. without a boundary) these integrals of motion follow
from the continuity equations for an infinite set of local currents. These
currents can be shown to exist in many 2D quantum field theories,
both the ones defined in terms of an action functional (like sine-Gordon
or nonlinear sigma-models[4]) and those which are defined as ``perturbed
conformal field theories''[5]. In presence of a boundary, the existense of
these currents is not sufficient to ensure integrability. Integrals of
motion appear only if particular ``integrable'' boundary conditions are
chosen. In general, the boundary condition can be specified either
through the
``boundary action functional'' or as the ``perturbed conformal boundary
condition''\footnote{$^2$}{Conformal field theories with boundary
are studied in [6,7,8] where, in particular, it is shown how to classify
conformally-invariant boundary conditions and boundary operators.}. In
Sect.2 we show how the integrals of motion of the ``bulk'' theory get modified
(in most cases destroyed) in presence of the boundary and how one can find
``integrable'' boundary conditions.

An important characteristic of an integrable field theory is its factorizable
S-matrix. In the ``bulk theory'', the factorizable S-matrix is completely
determined in terms of the two-particle scattering amplitudes, the latter
being required to satisfy the Yang-Baxter equation (also known as the
``factorizability condition''), in addition to the standard equations of
unitarity and crossing symmetry[4]. These equations have much restrictive
power, determining the S-matrix up to the so-called ``CDD ambiguity''.
At present many examples of factorizable scattering theory are known
(see e.g.[4,9-12]), most of which are obtained by explicitely solving the
above equations (eliminating the ``CDD ambiguity'' usually requires
a lot of guesswork).

It is known since long[13] that the concept of
factorizable scattering can be generalized in a  rather straightforward
way to the case where a reflecting boundary is present. The S-matrix
is expressed in terms of the ``bulk'' two-particle S-matrix and
specific ``boundary reflection'' amplitudes, the latter, again,
being required to satisfy an appropriate generalization of the Yang-Baxter
equation (which we call here the ``Boundary Yang-Baxter equation'')
\footnote{$^3$}{This equation found important applications in
quantum inverse scattering method[14], generalized to the systems with
boundary[15,16].}.  Generalization of the unitarity condition is also fairly
straightforward. What was not known was the appropriate analog of the
crossing-symmetry equation. In Section 3 we fill this gap by deriving
what we call the ``boundary cross-unitarity equation''. Together with this, the
above equations have exactly the same restrictive power as the
corresponding ``bulk'' system, i.e. they allow one to pin down the
factorizable boundary S-matrix up to the ``CDD factors''.

We also study integrable boundary conditions in two particular models
(off-critical Ising field theory and sine-Gordon theory) and find the
associated boundary S-matrices. This is done in Sections 4 and 5.

\vskip 0.4in

2.INTEGRALS OF MOTION

Consider a 2D Euclidean field theory, in flat space with coordinates
$(x^1,x^2)=(x,y)$.  There are basically two ways to define a 2D field
theory. In the Lagrangian approach one specifies the action

$${\cal A}=\int_{-\infty}^{\infty} dx \int_{-\infty}^{\infty} dy\quad
a(\varphi,\partial_\mu\varphi) \eqno (2.1)$$
where $\varphi(x,y)$ is some set of `` fundamental fields'' and the
action density $a(\varphi,\partial_{\mu}\varphi)$ is a local function of
these fields and derivatives $\partial_{\mu}\varphi= \partial\varphi /
\partial x^{\mu}$ with $\mu = 1,2$. Another approach is to consider the
``perturbed conformal field theory''; in this case one writes the ``symbolic
action''

$${\cal A} = {\cal A}_{CFT} +
\int_{-\infty}^{\infty}\Phi(x,y) dx dy \eqno (2.2)$$
where $A_{CFT}$ is the ``action of conformal field theory (CFT)'' and
$\Phi(x,y)$ is a specific relevant field of this CFT. In both approaches
one can define a symmetric stress tensor $T_{\mu\nu} = T_{\nu\mu}$ which
satisfies the continuity equations

$$\partial_{\bar z} T =
\partial_{z} \Theta\quad;\quad  \partial_{z} \bar T =\partial_{\bar z
} \Theta \eqno (2.3)$$
Here we use complex coordinates $z = x + iy$, $\bar z = x - iy$ and
denote the appropriate components $T = T_{zz}$, $\bar T = T_{\bar z\bar
z}$, $\Theta = T_{z\bar z}$ of the stress tensor. To achieve Hamiltonian
formulation one chooses an arbitrary direction, say the $y$-direction,
to be the ``euclidean time'', and associates a Hilbert space $\cal H$
with any ``equal time section'' $y = const.$, $x \in (-\infty, \infty)$.
States are vectors in $\cal H$ and their ``time evolution'' is described
 by the Hamiltonian operator

$$H = \int_{-\infty}^{\infty} dx T_{yy} = \int_{-\infty}^{\infty} dx [T + \bar
T + 2\Theta] \eqno (2.4)$$

Let us assume that the field theory (2.1) or (2.2) is integrable. In
particular, the equations (2.3) appear to be the first representatives
of an infinite sequence

$$\partial_{\bar z} T_{s+1} = \partial_{z} \Theta_{s-1} \quad \quad
\partial_{\bar z} \bar T_{s+1} = \partial_{z} {\bar\Theta}_{s-1}
\eqno       (2.5)$$
where $T_{s+1}, \Theta_{s-1}$ ($\bar T_{s+1}, \bar \Theta_{s-1}$)
are local fields of spins $s+1, s-1$ respectively and
the integrals of motion (IM)

$$P_{s} = \int_{-\infty}^{\infty}(T_{s+1} + \Theta_{s-1} ) dx;
\qquad \qquad \bar P_{s} = \int_{-\infty}^{\infty}
(\bar T_{s+1} + \bar \Theta_{s-1})\, dx \eqno (2.6)$$
constitute an infinite set of mutually commutative operators in $\cal
H$. The spin $s$ of IM (2.6) takes integer values $s_{1}, s_{2}, ...$
in the infinite set $\{s\}$ which is an important characteristic of
an integrable field theory[5]. In any case, $s_{1}=1$; for this value of
$s$ (2.5) coincides with (2.3) and

$$H =  (P_{1} + \bar P_{1}) \eqno (2.7)$$

Now, let us consider this field theory in the semi-infinite plane,
$x \in (-\infty,0], \quad y \in (-\infty,\infty)$, the $y$-axis
being the boundary. Again, the boundary conditions are specified in
different ways in the two approaches, (2.1) and (2.2). In the
lagrangian approach one chooses the ``boundary action density''
$b(\varphi_{B}(y),{d\over dy}\varphi_{B}(y))$, as a local function of
the ``boundary field'' $\varphi_{B},\quad \varphi_{B}(y) = \varphi(x,y)
|_{x=0}$, and writes the full action in the form

$${\cal A_{B}}=\int_{-\infty}^{\infty} dy \int_{-\infty}^{0} dx
a(\varphi, \partial_{\mu}\varphi) + \int_{-\infty}^{\infty} dy\;
b(\varphi_{B},{d\over dy}\varphi_{B}) \eqno (2.8)$$
To write down the analog of (2.2) in  presence of the boundary,
one starts with conformal field theory on the same semi-infinite plane
with certain conformal boundary conditions (CBC) at the boundary $x=0$,
and defines ``CBC perturbed by relevant boundary operator
$\Phi_{B}(y)$''. In general, this perturbation of boundary condition
goes along with the perturbation (2.2) of the bulk theory. This strategy
is summarised by the symbolic ``action''

$${\cal A} = {\cal A_{CFT+CBC}}\quad + \quad\int_{-\infty}^{\infty} dy
\int_{-\infty}^{0}dx\Phi(x,y) + \int_{-\infty}^{\infty}\,
dy\Phi_{B}(y) \eqno(2.9)$$

As is argued by Cardy[6], in CFT the equation $T_{xy}|_{x=0}=0$ is
satisfied with any choice of CBC. In the perturbed theory (2.9) this
condition is changed to

$$T_{xy}|_{x=0}=(-i)(T - \bar T)|_{x=0} = {d\over dy}\theta(y) \eqno
(2.10)$$
where $\theta(y)$ is some local boundary field. As the theory (2.8) or
(2.9) is still symmetric with respect to translations along the
$y$-axis, the equation (2.10) can be easily derived as a consequence of
this symmetry. In the ``perturbed CFT'' approach, it is relatively easy
to relate the field $\theta(y)$ to the ``boundary perturbation''
$\Phi_{B}(y)$(see below).

The continuity equations (2.3) guarantee that the contour
integrals

$$P_{1}({\cal C})=\int_{\cal C} (Tdz + \Theta d\bar z)\qquad; \qquad
\bar P_{1}({\cal C})=\int_{\cal C} (\bar T d\bar z + \Theta dz) \eqno (2.11)$$
do not change under deformations of the integration contours $\cal C$.
Consider the contour ${\cal C} = {\cal C}_{1} + {\cal C}_{12} + {\cal C}_{2}$
shown in Fig.1. Obviously, $P_{1}({\cal C})=\bar P_{1}({\cal C})=0$. If
we take the combination

$$0=P_{1}({\cal C}) + \bar P_{1}({\cal C}) = P_{1}(C_{1}) + \bar
P_{1}({\cal C}_{1}) + P_{1}({\cal C}_{2}) + \bar P_{1}({\cal C}_{2}) +
P_{1}({\cal C}_{12}) + \bar P_{1}({\cal C}_{12}) \eqno (2.12)$$
the integration over ${\cal C}_{12}$ part of this contour is
easily done in view of (2.10)

$$P_{1}({\cal C}_{12})+\bar P_{1}({\cal C}_{12}) = \theta(y_{1}) -
\theta(y_{2}) \eqno (2.13)$$
and hence the integral

$$H_{B}(y) = \int_{-\infty}^{0}(T + \bar T+ 2\Theta)\,dx\quad
+ \quad \theta(y) \eqno (2.14)$$
is in fact $y$-independent, i.e. it is an integral of motion in (2.8)
or (2.9).

Even if the bulk theory (2.1) or (2.2) is integrable, in general the
boundary conditions in the semi-infinite system (2.8) or
(2.9) will spoil integrability. Suppose, however, that we can choose
particular boundary conditions, such that the equation

$$[T_{s+1} + \bar \Theta_{s-1} - \bar T_{s+1} - \Theta_{s-1}]|_{x=0} =
{d\over dy} \theta_{s}(y) \eqno (2.15)$$
is satisfied for some $s \in \{s\}$; here again $\theta_{s}(y)$ is
some local boundary field. Then repeating the above argument, one finds
that the quantity

$$H_{B}^{(s)}=\int_{-\infty}^{0}[T_{s+1}(x,y)+\Theta_{s-1}(x,y)+\bar
T_{s+1}(x,y)+\bar\Theta_{s-1}(x,y)]\,dx\quad +\quad \theta_{s}(y) \eqno
(2.16)$$
does not depend on $y$, i.e. it appears as a non-trivial integral of
motion. We will call the boundary conditions ``integrable'' (and refer
to the theory (2.8)((2.9)) as ``integrable boundary field theory'') if
the equation (2.15) holds for any $s$ out of an infinite set $\{s\}_B$,
subset in $\{s\}$.

There are two alternative natural ways to introduce the
Hamiltonian picture in the theory (2.8)((2.9)). First, one can take
again the direction along the boundary ($y$-direction) to be the ``time''.In
this case the boundary appears as the ``boundary in space'', and
the Hilbert space of states ${\cal H}_{B}$ is associated with the
semi-infinite line $y=const$, $x \in (-\infty,0]$. Then the
quantities (2.14), (2.16) appear as operators acting in ${\cal
H}_{B}$, and $H_{B}(=H_{B}^{(1)})$ is naturally identified with
the Hamiltonian. The correlation functions of any local fields
$O_{i}(x,y)$ in presence of the boundary can be computed in this
picture as the matrix elements

$$\langle O_{1}(x_{1},y_{1}) ... O_{N}(x_{N},y_{N})\rangle =
{{_{B}\langle 0 \mid {\cal T}_{y}O_{1}(x_{1},y_{1}) ...
O_{N}(x_{N},y_{N})\mid 0 \rangle_{B}} \over {_{B}\langle 0 \mid
0 \rangle_{B}}} \eqno (2.17)$$
where $\mid 0 \rangle_{B}$$ \in {\cal H}_{B}$ is the ground state of
$H_{B}$, and $O_{i}(x_{i},y_{i})$ in the r.h.s. are understood as
the corresponding Heisenburg operators

$$O_{i}(x,y)=e^{-yH_{B}}O_{i}(x,0)e^{yH_{B}} \eqno (2.18)$$
and ${\cal T}_{y}$ means the ``$y$-ordering''. In an integrable theory the
operators $H_{B}^{(s)};\quad s \in \{s\}_B$, constitute a commutative
set of IMs.

Alternatively, one could take $x$ to be the ``euclidean
time''. In this case the ``equal time section'' is the infinite line
$x=const$, $y \in (-\infty,\infty)$. Hence the associated space of
states is the same $\cal H$ as in the bulk theory (2.1)((2.2)), and
the Hamiltonian operator is given by the same eq.(2.4) (with $x$ and
$y$ interchanged). The boundary at $x=0$ appears as the ``time
boundary'', or ``initial condition'' at $x=0$ which is described by
the particular ``boundary state''\footnote{$^4$}{The notion of boundary
state is discussed in the context of CFT in [7].}$\mid B\rangle \in
{\cal H}$. It is the state $\mid B\rangle$ that concentrates all
information about the boundary condition in this picture. The
correlation functions (2.17) are expressed as

$$\langle O_{1}(x_{1},y_{1}) ... O_{N}(x_{N},y_{N})\rangle =
{\langle 0 \mid {\cal T}_{x}(O_{1}(x_{1},y_{1}) ... O_{N}(x_{N},y_{N})
\mid B  \rangle \over \langle 0\mid B \rangle}  \eqno (2.19)$$
where now $\mid 0\rangle \in {\cal H}$ is the ground state of $H$ and
$O_{i}(x,y)$ in the r.h.s. are the Heisenberg field operators

$$O_{i}(x,y)=e^{-xH}O_{i}(0,y)e^{xH} \eqno (2.20)$$
corresponding to this picture; ${\cal T}_{x}$ means ``$x$-ordering''.
In an integrable theory, the same equations (2.6) (again with $x$ and $y$
interchanged) define an infinite set of mutually commutative operators
$P_{s},\bar P_{s};s \in \{s\}$, acting in $\cal H$. As a
direct consequence of (2.15) one finds that the boundary state $\mid B
\rangle$ satisfies the equations

$$(P_{s}-\bar P_{s})\mid B\rangle = 0 \quad ;\quad
s \in \{s\}_B . \eqno (2.21)$$

To expose an example of an integrable boundary field theory, let us cosider
the perturbed CFT (2.9) with ${\cal A}_{CFT}$ taken to be any $c<1$
minimal model, and the degenerate spinless field $\Phi_{(1,3)}$ taken as
the bulk perturbation,

$$\Phi_{(x,y)}=\lambda\Phi_{(1,3)}(x,y). \eqno (2.22)$$
where $\lambda$ is a constant of dimension ${\hbox
{[length]}}^{2\Delta-2}; \Delta =\Delta_{ (1,3)}$. The local integrals
of motion in the corresponding bulk theory (2.2) are discussed in [5].
The fields $T_{s+1}$ are composite fields built up from $T=T_{zz}$ and
its derivatives, for example

$$T_{2}=T\quad ;\quad T_{4}=:T^{2}: \quad;\quad
T_{6}=:T^{3}:-{{c+2}\over 6}:(\partial_{z} T)^{2}:\quad ;
\quad ... \eqno (2.23)$$
where :  : denotes appropriately regularized products ($\bar
T_{s+1}$ are built from $\bar T$ in similar ways). The
characteristic feature of the fields $T_{s+1}$ is that, in the
conformal limit $\lambda=0$ their OPE's with $\Phi_{(1,3)}$ have the
form

$$T_{s+1}(z)\Phi_{(1,3)}(w, \bar w)=\sum_{k=1}^{s}{\Psi_{s-k+1}^{(k)}(w,
\bar w)\over (z-w)^{k}}\quad +\quad \hbox {regular terms} \eqno (2.24)$$
($\Psi_{s-k+1}^{(k)}$ are particular conformal descendants of
$\Phi_{(1,3)}$) with

$$\Psi_{s}^{(1)}(w, \bar w)=\partial_{w}Q_{s-1}(w,\bar w), \eqno
(2.25)$$
where  $Q_{s-1}$ are local fields. The ``null-vector'' equation

$$(L_{-3}-{2\over{\Delta + 1}}L_{-1}L_{-2} + {1\over{(\Delta + 1)(\Delta
+ 2)}}L_{-1}^{3})\Phi_{(1,3)}, \eqno(2.26)$$
satisfied by the degenerate field $\Phi_{(1,3)}$, is used to prove
(2.25). The equation (2.25) is sufficient to show that the field $T_{s+1}$
satisfy (2.5) in the perturbed theory, up to first order in $\lambda$.
Then dimensional analysis shows that (2.5) is exact. There are
infinitely many fields $T_{s+1}$ satisfying (2.24)-(2.25), one for each
odd $s$. So, the set $\{s\}$ in this theory contains all positive odd
integers.

Conformal boundary conditions in $c=1-{6\over p(p+1)}$ minimal CFT are
classified in [7]. There are finitely many possible conformal boundary
conditions, each corresponding to a particular cell $(r,s)$ of the Kac
table. In each case possible boundary operators are degenerate primary
boundary fields $\psi_{(n,m)}$ (plus their Virasoro descendants), such
that the fusion rule coefficients $N_{(n,m)(r,s)}^{(r,s)}$ are non-zero
(the other degenerate boundary fields, with $N_{(n,m)(r,s)}^{(r,s)}=0$
correspond to ``juxtapositions'' of different conformal boundaries,
see [7] for details). The field $\Phi_{(1,3)}$ satisfies this condition
for any $(r,s)$. We take this field to be the boundary perturbation
in (2.9), i.e.

$$\Phi_{B}(y)=\lambda_{B}\psi_{(1,3)}(y). \eqno (2.27)$$
Its conformal dimension $\Delta = \Delta_{(1,3)} < 1$, so that it is a
relevant perturbation. We want to show that under this choice the
theory (2.9) is integrable.

To warm up, let us consider the components $T$ and $\bar T$ of the
stress tensor itself. In the conformal limit $\lambda =0,\lambda_{B}=0$,
these components satisfy the boundary condition

$$[T(y+ix)-\bar T(y-ix)]|_{x=0}=0,\eqno (2.28)$$
i.e. the field $\bar T(\bar z)$ is just the analytic continuation
of $T(z)$ to the lower half-plane $Im z < 0$. Let us ``turn on'' the
boundary perturbation, $\lambda_{B}{\not =}0$, still keeping $\lambda
=0$, and consider the correlation function

$$\langle [T(y+ix)-\bar T(y-ix)]X\rangle_{\lambda_{B}} = $$
$$Z_{\lambda_{B}}^{-1} \langle [T(y+ix)-\bar T(y-ix)]X e^{-{\lambda_{B}}
\int_{-\infty}^{\infty}\psi_{(1,3)}(y')dy'}\rangle_{CFT},
\eqno (2.29)$$
where $X$ is any product of fields located away from the
boundary $x=0$ and $Z_{\lambda_{B}}^{-1}=\langle
e^{-\lambda_{B}\int_{-\infty}^{\infty}\psi_{(1,3)}(y)dy}\rangle_{CFT}$.
In the limit $x\to 0$ the contribution to (2.29) is controlled by OPE

$$(T(y+ix)-T(y-ix))\lambda_{B}\psi_{(1,3)}(y')=$$
$$\lambda_{B}\{{\Delta_{(1,3)}\over
(y-y'+ix)^{2}}-{\Delta_{(1,3)}\over (y-y'-ix)}+{1\over
(y-y'+ix)}{\partial\over \partial y'}-{1\over
(y-y'-ix)}{\partial\over \partial y'}\}\psi_{(1,3)}(y') \to$$ $$\to
\lambda_{B}\{\Delta_{(1,3)}\delta'(y-y')+\delta(y-y'){\partial\over
\partial y'}\}\psi_{(1,3)}(y'). \eqno (2.30)$$
This shows that for $\lambda_{B}\neq 0$ the fields $T(x,y)$ and $\bar T(x,y)$
satisfy (2.10) with
$$\theta (y)=(1-\Delta)\lambda_{B}\psi_{(1,3)}(y). \eqno (2.31)$$
Now, dimensional analysis, exactly parallel to that carried out in [5]
shows that the equations (2.10),(2.30) remain valid if we turn on the
bulk perturbation, i.e. at $\lambda \neq 0$.

The above arguments can be repeated for the higher
currents $T_{s+1}$, $s=3, 5, ...$. To see this, note that the null
vector equation (2.26), crucial for the validity of (2.25), is satisfied by
$\psi_{(1,3)}$ as well. Therefore, in the conformal limit
$\lambda=\lambda_{B}=0$ the fields $T_{s+1}(z)$ satisfy the OPEs

$$T_{s+1}(z)\psi_{(1,3)}(y)=\sum_{k=1}^{s}{1\over
(z-y)^{k}}\chi_{s+1-k}^{(k)}(y)\quad +\quad \hbox {regular
terms},$$
$$\chi_{s}^{(1)}(y)={d\over dy}q_{s}(y), \eqno(2.32)$$
similar to (2.24),(2.25) where $q_{s},\chi$ are descendants of
$\psi_{(1,3)}$. Whence

$$\hbox {limit}_{x\to 0}(T_{s+1}(y+ix)-T_{s+1}(y-ix))\,
\lambda_{B}\psi_{(1,3)}(y')$$
$$=\lambda_{B}(\delta(y-y'){d\over
dy'}q_{s}(y')+\sum_{k=2}^{s}{1\over{(k-1)!}}{d^{k-1}\over dy'^{k-1}}
\delta(y-y')\chi_{s+1-k}^{k}(y')), \eqno (2.33)$$
and we conclude that for $\lambda =0$ and in the first order in
$\lambda_{B}$ the equation holds

$$[T_{s+1}(x,y)-\bar T_{s+1}(x,y)]_{x=0}={d\over
dy}\theta_{s} \eqno (2.34)$$
with

$$\theta_{s}(y)=\lambda_{B}[q_{s}(y)+\sum_{k=2}^{s}{1\over{(k-1)!}}{d^{k-1}
\over dy^{k-1}}\chi_{s+1-k}^{k}(y)]. \eqno (2.35)$$
The higher powers of $\lambda_B$ can contribute to (2.34) through
the ``resonance terms'' similar to those discussed in [5]. It is plausible,
however,that these do not spoil the general form of (2.34) but simply
modify (2.35) by higher order terms in $\lambda_{B}$.
It is also plausible (and can be supported to some extent by dimensional
analysis of [5]) that ``turning on'' the bulk perturbation, $\lambda \neq
0$, converts (2.34) to(2.15).

We realise that the above arguments do not constitute a rigorous proof.
First, we did not solve the problem of the ``resonance terms'' (this
problem remains open in the bulk theory, too). More importantly,
we did not analyse possible effects of mixing between the boundary
and the bulk perturbations. We have shown,however,that the field theory
(2.9) with (2.22) and (2.26) satisfies some very non-trivial necessary
(but not sufficient) conditions of integrability and we conjecture that
this boundary field theory with boundary is integrable.

\vskip 0.4in

3.BOUNDARY S-MATRIX

If the field theory (2.1) ((2.2)) is massive the space $\cal H$ is
the Fock space
of multiparticle states. After rotation $y=it$ to $1+1$ Minkowski
space-time these states are interpreted as the asymptotic (``in-'' or
``out-'') scattering states. For an integrable field theory the scattering
is purely elastic and the corresponding S-matrix is factorizable. The
factorizable scattering theory in infinite space is discussed in many
papers and reviews (see e.g.[4,5,9-12]); below, we describe just some
basics of the theory. The boundary theory (2.8)((2.9)) in Minkowski
space is also interpreted as a scattering theory. For the integrable
boundary field theory
this scattering theory is  again purely elastic and the corresponding
S-matrix is the ``factorizable boundary S-matrix''. The ``factorizable
boundary scattering theory'' is developed in close parallel with the
``bulk'' theory (see [13]). However there are still some gaps in this
parallel (most importantly, the boundary analog of crossing-symmetry
condition of the ``bulk'' S-matrix is not absolutely straightforward).
It is the aim of this Section to fill these gaps.

We start with a brief description of the basics of the factorizable
scattering theory in the infinite space (``bulk theory''). Assume that
the theory contains $n$ sorts of particles $A_a$; $a=1,2,...,n$ with the
masses $m_a$. As usual, we describe the kinematic states of the
particles in terms of their rapidities $\theta$,

$$p_0 + p_1 = me^{\theta}; \qquad  p_0 - p_1 = me^{-\theta},   \eqno(3.1)$$
where $p_\mu$ are the components of the two-momentum and $m$ is the particle
mass. The asymtotic particle states are generated by the ``particle creation
operators'' $A_{a}(\theta)$

$$\mid A_{a_1} (\theta_1) A_{a_2} (\theta_2) ... A_{a_N} (\theta_N)\rangle =
 A_{a_1} (\theta_1) A_{a_2} (\theta_2) ... A_{a_N} (\theta_N)\mid 0 \rangle.
\eqno(3.2)$$
The state (3.2) is interpreted as an ``in-state'' if the rapidities
$\theta_i$ are ordered as $\theta_1 > \theta_2 > ... > \theta_N$; if
instead $\theta_1 < \theta_2 < ... < \theta_N$  (3.2) is understood as an
``out-state'' of scattering. The ``creation operators'' $A_a(\theta)$
satisfy the commutation relations

$$A_{a_1}(\theta_1)A_{a_2}(\theta_2)=S_{a_1 a_2}^{b_1 b_2}(\theta_1 -
\theta_2)A_{b_2}(\theta_2)A_{b_1}(\theta_1) \eqno(3.3)$$
which are used to relate the ``in-'' and the ``out-'' bases and hence
completely
describe the S-matrix. The coefficient functions $S_{a_1 a_2}^{b_1
b_2}(\theta)$ are interpreted as the two-particle scattering amplitudes
describing the processes $A_{a_1}A_{a_2} \to  A_{b_1}A_{b_2}$ (see
Fig.2). The asymptotic states (3.2) diagonalise the local IM (2.6), the
eigenvalues being determined by the relations

$$[P_s,A_a(\theta)] = \gamma_{a}^{(s)}e^{s\theta}A_a (\theta);
\qquad [\bar P_s,A_a(\theta)] = \gamma_{a}^{(s)}e^{-s\theta}A_a (\theta);
\eqno(3.4)$$

$$P_s\mid 0 \rangle =0; \qquad \bar P_s\mid 0 \rangle =0, \eqno(3.5)$$
where $\gamma_{a}^{(s)}$ are constants ($\gamma_{a}^{(1)}=m_a$). These
IM must commute with the S-matrix; it follows in particular that the
amplitude $S_{a_1 a_2}^{b_1 b_2}(\theta)$ is zero unless
$m_{a_1}=m_{b_1}$ and $m_{a_2}=m_{b_2}$ (other concequences are
discussed in [5]).

Charge conjugation $\bf C$ acts as an involution of the set of particles
$\lbrace A_a \rbrace$, i.e. ${\bf C}: A_a \leftrightarrow A_{\bar a}$,
where $A_{\bar a}\in $$\lbrace A_a \rbrace$, so that each particle in
$\lbrace A_a \rbrace$ is either neutral ${\bf C}A_a =A_a$ or belongs to the
particle-antiparticle pair $(A_a, A_{\bar a})$. In this Section we
assume for simplicity that the theory under consideration respects $\bf
C$, $\bf P$ and $\bf T$ symmetries, i.e.

$$S_{a_1 a_2}^{b_1 b_2}(\theta) = S_{\bar a_1 \bar a_2}^{\bar b_1
\bar b_2}(\theta) = S_{a_2 a_1}^{b_2 b_1}(\theta) =  S_{\bar b_2 \bar
b_1}^{\bar a_2\bar a_1}(\theta). \eqno(3.6)$$

The two-particle S-matrix $S_{a_1 a_2}^{b_1 b_2}(\theta)$ is the basic
object of the theory. It must satisfy several general requirements

1.Yang-Baxter (or ``factorization'') equation

$$S_{a_1 a_2}^{c_1 c_2}(\theta)S_{c_1 a_3}^{b_1
c_3}(\theta+\theta')S_{c_2 c_3}^{b_2 b_3}(\theta') = S_{a_2 a_3}^{c_2
c_3}(\theta')S_{a_1 c_3}^{c_1 b_3}(\theta+\theta') S_{c_1 c_2}^{b_1
b_2}(\theta); \eqno(3.7)$$
here and below summation over repeated indices is assumed. This equation
is illustrated in Fig.3. Formally, this equation appears as the associativity
condition for the algebra (3.3).

2.Unitarity condition

$$S_{a_1 a_2}^{c_1 c_2}(\theta)S_{c_1 c_2}^{b_1 b_2}(-\theta) =
\delta_{a_1}^{b_1} \delta_{a_2}^{b_2}. \eqno(3.8)$$
Graphic representation of this equation is shown in Fig.4. It can also
be obtained as the consistancy condition for the algebra (3.3) (one applies
(3.3) twice).

3.Analyticity and Crossing symmetry. The amplitudes $S_{a_1 a_2}^{b_1
b_2}(\theta)$ are meromorphic functions of $\theta$, real at $Re\theta =
0$. The domain $0 < Im\theta < \pi$ is called the ``physical strip''.
The physical scattering amplitudes of the ``direct channel'' $A_{a_1}A_{a_2}
\to  A_{b_1}A_{b_2}$ are given by the values of the functions $S_{a_1
a_2}^{b_1 b_2}(\theta)$ at $Im\theta = 0$, $Re\theta > 0$. The values of
these functions at $Im\theta = 0$, $Re\theta < 0$ describe the amplitudes
of the
``cross-channel''  $A_{a_2}A_{\bar b_1} \to  A_{b_2}A_{\bar a_1}$. The
functions $S_{a_1 a_2}^{b_1 b_2}(\theta)$ satisfy the crossing symmetry
relation

$$S_{a_1 a_2}^{b_1 b_2}(\theta) = S_{a_2 \bar b_1}^{b_2 \bar a_1}(i\pi -
\theta) \eqno(3.9)$$
(see Fig.5). Combining this and the Eq.(3.8) one can derive the
``cross-unitarity equation''

$$S_{a_1 c_2}^{c_1 b_2}(i\pi - \theta)S_{a_2 c_1}^{c_2 b_1}(i\pi +
\theta) = \delta_{a_1}^{b_1}\delta_{a_2}^{b_2}. \eqno(3.10)$$

4.Bootstrap condition. The only singularities of $S_{a_1 a_2}^{b_1
b_2}(\theta)$ admitted in the physical strip are poles located at
$Re\theta = 0$. The simple poles are interpreted as bound states, either
of the direct or of the cross channel. As the bound states are stable
particles they must be in the set $\lbrace A_a \rbrace$. Let $iu_{a_1
a_2}^{c}$ be the position of the pole of $S_{a_1 a_2}^{b_1 b_2}(\theta)$
associated with the ``bound state'' $A_c$ of the direct channel. Then
$u_{a_1 a_2}^{c}$ must satisfy the relation

$$m_{a_1}^2 + m_{a_2}^2 - m_c^2 = -2m_{a_1}m_{a_2}\cos u_{a_1 a_2}^{c},
\eqno(3.11)$$
i.e. the quantity $\bar  u_{a_1 a_2}^{c} = \pi - u_{a_1 a_2}^{c}$ can be
interpreted as the  internal angle of the euclidean triangle with the sides
$m_{a_1}, m_{a_2}, m_c$. The pole term

$$S_{a_1 a_2}^{b_1 b_2}(\theta) \simeq i\,{{f_{a_1 a_2}^c f_{c}^{b_1
b_2}}\over {\theta - iu_{a_1 a_2}^{c}}} \eqno(3.12)$$
corresponds to the diagram in Fig.6, where the vertices represent
the ``three-particle couplings'' $f$ (Fig.7). In this situation the
two-particle S-matrix satisfies the ``bootstrap equation''

$$f_{a_1 a_2}^{c}S_{c a_3}^{b b_3}(\theta)= f_{c_1 c_2}^{b}S_{a_1
c_3}^{c_1 b_3}(\theta + i\bar u_{a_1 \bar c}^{\bar a_2})S_{a_2 a_3}^{c_2
c_3}(\theta - i\bar u_{a_2 \bar c}^{\bar a_1}) \eqno(3.13)$$
which is illustrated by Fig.8.

More details about the factorizable scattering theory and many examples can
be found in the original papers and reviews (see e.g.[4,5,9-12]). Most of the
examples are constructed directly, by solving the Eq.(3.7-3.9) above.
Following this approach one can pin down the S-matrix $S_{a_1 a_2}^{b_1
b_2}(\theta)$ up to the so-called ``CDD ambiguity''

$$S_{a_1 a_2}^{b_1 b_2}(\theta) \to S_{a_1 a_2}^{b_1 b_2}(\theta)\Phi
(\theta), \eqno(3.14)$$
where the ``CDD factor'' $\Phi (\theta)$ is an arbitrary function
satisfying the equations

$$\Phi(\theta) = \Phi (i\pi - \theta); \qquad \Phi (\theta)\Phi(-\theta)
= 1. \eqno(3.15)$$
The bootstrap equation may impose further restrictions on this function.

Let us turn now to the semi-infinite system (2.8)((2.9)). Here again the
states in ${\cal H}_B$ can be classified as asymptotic scattering
states. The scattering occurs in the semi-infinite $1+1$ Minkowski
space-time $(x,t), t=-iy, x<0$. The initial state

$$\mid A_{a_1}(\theta_1)A_{a_2}(\theta_2)...
A_{a_N}(\theta_N)\rangle_{B, in} \eqno(3.16)$$
(the subscript $B$ indicates that $\mid ...\rangle_B \in {\cal H}_B$) of
the scattering consists of some number ($N$) of ``incoming'' particles
moving towards the boundary at $x=0$, i.e. all the rapidities $\theta_1,
\theta_2,...,\theta_N$ are positive. In the infinite future, $t \to
\infty$, this state becomes a superposition of the ``out-states''

$$\mid A_{b_1}({\theta_1}')A_{b_2}({\theta_2}')...
A_{b_M}({\theta_M}')\rangle_{B, out} \eqno(3.17)$$
each containing some number of ``outgoing'' particles moving away from
the boundary with negative rapidities
${\theta_1}',{\theta_2}',...,{\theta_M}'$. In integrable boundary field
theory this process is constrained by the IM (2.16). Like in the
``bulk'' theory the operators $H_s$ are diagonal in the basis of
asymptotic states and

$$H_s\mid
A_{a_1}(\theta_1)A_{a_2}(\theta_2)...A_{a_N}(\theta_N)\rangle_{B,
in(out)} = $$
$$(\sum_{i=1}^{N}2\gamma_{a_i}^{(s)}\cosh(s\theta_i) + h^{(s)})\mid
A_{a_1}(\theta_1)A_{a_2}(\theta_2)...A_{a_N}(\theta_N)\rangle_{B, in(out)}
\eqno(3.18)$$
where $h^{(s)}$ are some constants. The constraints

$$\sum_{i=1}^{N}\gamma_{a_i}^{(s)}\cosh(s\theta_i) =
\sum_{j=1}^{M}\gamma_{b_j}^{(s)}\cosh(s{\theta_j}') \eqno(3.19)$$
which follow from (3.18) show that $M=N$ and the set of rapidities
$\lbrace {\theta_1}',{\theta_2}',...,{\theta_N}'\rbrace$ can differ only
by permutation from $\lbrace -\theta_1,-\theta_2,...,-\theta_N\rbrace$,
i.e. the boundary scattering theory is purely elastic. It is possible to
argue that the S-matrix in this case has a factorizable structure.

The factorizable boundary scattering theory can be described in complete
analogy with the ``bulk'' scattering theory. Again, the asymptotic states
(3.16),(3.17) are generated by the ``creation operators'' $A_a(\theta)$
satisfying the same commutation relations (3.3). If $\theta_1 > \theta_2
> ... > \theta_N > 0$ the in-state (3.16) can be written as

$$A_{a_1}(\theta_1)A_{a_2}(\theta_2)...A_{a_N}(\theta_N)\mid 0
\rangle_B, \eqno(3.20)$$
where $\mid 0 \rangle_B$ is the ground state of $H_B$. One can think of
the boundary as an infinitely heavy impenetrable particle $B$ sitting at
$x=0$ and formally write the state $\mid 0 \rangle_B$ as

$$\mid 0 \rangle_B = B\mid 0 \rangle \eqno(3.21)$$
in terms of ``operator'' $B$ which we call the ``boundary creating
operator''(formally $B: {\cal H} \to {\cal H}_B$; it is an interesting
question wether (3.21) makes any more than just formal sense).
The ``operator'' $B$ satisfies the relations

$$A_a (\theta)B = R_{a}^{b}(\theta)A_b (-\theta)B, \eqno(3.22)$$
the coefficient functions $R_{a}^{b}(\theta)$ being interpreted as the
amplitudes of one-particle reflection off the boundary, as shown in
Fig.9. The Eq.(3.18) is reproduced if we assume (3.4), (3.5) and
$[H_s,B]= h^{(s)} B$. It follows from (3.17) that $R_{a}^{b}(\theta)$
vanishes if $m_a \neq m_b$. By purely algebraic manipulations, with the
use of relations (3.3) and (3.22), one can expand any in-state (3.20) in
terms of the out-states

$$ A_{b_1}(-\theta_1)A_{b_2}(-\theta_2)...A_{b_N}(-\theta_N)\mid 0
\rangle_B, \eqno(3.23)$$
($\theta_1 > \theta_2 > ... > \theta_N > 0$) thus expressing the
$N$-particle S-matrix

$$A_{a_1}(\theta_1)A_{a_2}(\theta_2)...A_{a_N}(\theta_N)\mid 0\rangle_B
=$$
$$R_{a_1 a_2 ... a_N}^{b_1 b_2 ... b_N}(\theta_1, \theta_2, ...,
\theta_N)A_{b_1}(-\theta_1)A_{b_2}(-\theta_2)...A_{b_N}(-\theta_N)\mid 0
\rangle_B \eqno(3.24)$$
in terms of the ``fundamental amplitudes'' $S_{a_1 a_2}^{b_1
b_2}(\theta)$ and $R_{a}^{b}(\theta)$ which are basic objects of
the factorizable boundary scattering theory. The amplitudes
$R_{a}^{b}(\theta)$ have to satisfy several general requirements
analogous to the requirements $1-4$ of the ``bulk'' theory above.

$1'$.Boundary Yang-Baxter equation

$$R_{a_2}^{c_2}(\theta_2)S_{a_1 c_2}^{c_1 d_2}(\theta_1 +
\theta_2)R_{c_1}^{d_1}(\theta_1)S_{d_2 d_1}^{b_2 b_1}(\theta_1 -
\theta_2) = $$
$$S_{a_1 a_2}^{c_1 c_2}(\theta_1 -
\theta_2)R_{c_1}^{d_1}(\theta_1)S_{c_2 d_1}^{d_2 b_1}(\theta_1 +
\theta_2)R_{d_2}^{b_2}(\theta_2) \eqno(3.25)$$
(represented graphically in Fig.10) can be obtained as the associativiy
condition of the algebra (3.22), (3.3). These equations have been
introduced first in [13] and studied in relation with the quantum inverse
scattering method for the integrable systems with boundary in many
subsequent papers (see e.g. [15-17]). Note that (3.25) is the direct
analog of (3.7).

$2'$.Boundary Unitarity condition

$$R_{a}^{c}(\theta)R_{c}^{b}(-\theta)=\delta_{a}^{b} \eqno(3.26)$$
is also an absolutely straightforward generalization of (3.8) (see Fig.11).
One obtains (3.26) applying (3.22) twice.

$3'$.The boundary analog of crossing-symmetry condition (3.9) is far less
straightforward. Note that without any additional conditions the
equations (3.25) and (3.26) are not restrictive enough. The ambiguity in
the solution is

$$R_{a}^{b}(\theta) \to R_{a}^{b}(\theta) \Phi_B (\theta) \eqno(3.27)$$
with arbitrary function $\Phi_B$ which satisfy the equation

$$\Phi_B (\theta)\Phi_B (-\theta) = 1 \eqno(3.28)$$
only, which does not even imply (contrary to (3.15)) that  $\Phi_B$ is
analytic (meromorphic) in the full complex plane of $\theta$.

To reveal the analog of the ``cross channel'' of the scattering process
(3.24) one has to use the alternative Hamiltonian picture for the
boundary field theory mentioned in Sect.2. Consider $1+1$ Minkowski
space-time $(\tau, y)$ with $\tau = ix (\tau > 0)$ interpreted as time.
The ``equal time'' section now is the infinite line $-\infty < y < \infty$
and the space of states $\cal H$ is the same as in the ``bulk'' theory.
The boundary condition at $x = 0$ appears in this picture as the
initial condition at $\tau = 0$; it is described by the ``boundary
state'' $\mid B \rangle$ as explained in Sect.2. As $\mid B \rangle \in
\cal H$ this state is a superposition of the asymptotic states (3.2) of
the ``bulk'' theory. In integrable theory with ``integrable boundary''
the states (3.2) admitted to contribute to $\mid B \rangle$ are
restricted to satisfy (2.21). Noting that the eigenvalue of the operator
$P_s - \bar P_s$ on (3.2) is

$$\sum_{i=1}^{N}2\gamma_{a_i}^{(s)}\sinh(s\theta) \eqno(3.29)$$
we conclude that the particles $A_a$ can enter the state $\mid B
\rangle$ only in pairs $A_a (\theta)A_b (-\theta)$ of equal-mass
particles of the opposite rapidities\footnote{$^5$}{The possibility
of zero rapidity particles $A_a (0)$ entering the boundary state, which is
evidently consistent with (2.21), is discussed below.}. Thus we can write

$$\mid B \rangle = {\cal N}\sum_{N=0}^{\infty}\int_{0 < \theta_1 <
\theta_2 < ...
< \theta_N}d\theta_1 d\theta_2 ... d\theta_N K_{2N}^{a_N a_{N-1} ...
a_1, b_1 b_2 ... b_N}(\theta_1, \theta_2, ..., \theta_N)$$
$$A_{a_N}(-\theta_N)A_{a_{N-1}}(-\theta_{N-1})...A_{a_1}(-\theta_1)A_{b_1}(\theta_1)A_{b_2}(\theta_2)...A_{b_N}(\theta_N)\mid 0 \rangle \eqno(3.30)$$
where we have chosen the expansion in terms of the out-states; $K_{2N}$ are
certain
coefficient functions which can be related to the amplitudes $R$ in
(3.24). The overall factor $\cal N$ in (3.30) is chosen in such a way that
$K_0=1$; it is possible to argue that in the massive theory
with a non-degenerate ground state $\mid 0 \rangle_B$, which we consider
here,  one can choose ${\cal N} =1$ by adding an appropriate
constant term to the boundary action density $b$ in (2.8) (or to the
``perturbing field'' $\phi_B$ in (2.9)); in what follows we assume this
choice. By applying the standard reduction technique to the equality
(2.17), (2.19) one can show that (under appropriate normalization of the
operators $A_{a}(\theta)$) the following equations hold

$$K^{a_N a_{N-1} ... a_1, b_1 b_2 ... b_N}(\theta_1, \theta_2, ...,
\theta_N)= R_{\bar a_1 \bar a_2 ... \bar a_N}^{b_1 b_2 ...
b_N}({i\pi\over 2} -\theta_1, {i\pi \over 2}-\theta_2, ...,
{i\pi \over 2}-\theta_N), \eqno(3.31)$$
up to a constant phase factor which depends on the normalizations of the
operators $A_{a}(\theta)$; in what follows we assume that they are
normalized in such a way that (3.31) holds as it stands.

Let us concentrate attention on the amplitude $K^{a b}(\theta)
\equiv K_{2}^{a,b}(\theta)$ describing the contribution

$$\mid B \rangle = (1 +
\int_{0}^{\infty}K^{a b}(\theta)A_a(-\theta)A_b(\theta) + ...)\mid 0
\rangle \eqno(3.32)$$
of the two-particle out-state $A_a(-\theta)A_b(\theta)\mid 0 \rangle$ to the
boundary state $\mid B \rangle$. It satisfies

$$K^{a b}(\theta)=R_{\bar a}^{b}({i\pi \over 2}-\theta) \eqno(3.33)$$
i.e. the ``elementary reflection amplitude''$R_{\bar a}^{b}(\theta)$ can
be obtained by analytic continuation of the amplitude $K^{a b}(\theta)$
to the domain $Im\theta = {i\pi \over 2}, Re\theta < 0$. The values of
$K^{a b}(\theta)$ at negative real $\theta$ (corresponding to the ``lower
edge'' of the cut in the energy plane) are interpreted as the
coefficients of expansion of $\mid B \rangle$ in terms of the $in-$states
$A_a(\theta)A_b(-\theta)\mid 0 \rangle, \theta > 0$

$$\mid B \rangle = (1 +
\int_{0}^{\infty}K^{a b}(-\theta)A_a(\theta)A_b(-\theta) + ...)\mid 0
\rangle \eqno(3.34)$$
As the $in-$ and the $out-$ states are related through the S-matrix, the
amplitude $K^{a b}(\theta)$ has to satisfy the following
``boundary cross-unitarity condition''

$$K^{a b}(\theta) = S_{a' b'}^{a b}(2\theta)K^{b' a'}(-\theta)
\eqno(3.35)$$
which we consider to be the boundary analog of (3.10). It is illustrated
by the diagram in Fig.12. With this equation added, the ambiguity in the
solution of (3.25), (3.26) and
(3.35) reduces to (3.27) with $\Phi_B$ satisfying (3.28) and

$$\Phi_B (\theta) = \Phi_B (i\pi - \theta) \eqno(3.36)$$
which is exactly the same as the ``CDD umbiguity'' (3.14), (3.15) in the
``bulk'' theory. Let us note that (3.35) allows one to write (3.32) as

$$\mid B \rangle = (1 + {1\over 2}
\int_{-\infty}^{\infty}K^{a b}(\theta)A_a(-\theta)A_b(\theta) + ...)\mid 0
\rangle \eqno(3.37)$$

Using the Equation (3.31) one can express all the amplitudes $K_{2N}$ in
(3.30) in terms of the two-particle boundary amplitudes $K^{a
b}(\theta)$ and the elements of the two-particle S-matrix $S_{a b}^{c
d}(\theta)$. The result is in exact agreement with the following
simple expression

$$\mid B \rangle = \Psi[K(\theta)]\mid 0 \rangle =
exp(\int_{-\infty}^{\infty}d\theta K(\theta))\mid 0\rangle \eqno(3.38)$$
where

$$K(\theta)={1\over 2}K^{a b}(\theta)A_a(-\theta)A_b(\theta)
\eqno(3.39)$$
Note that although the ``creation operators'' $A_a(\theta)$ do not
comute, the bilinear expressions (3.39) satisfy the commutativity conditions

$$[K(\theta), K(\theta')] = 0 \eqno(3.39)$$
as a direct consequence of the ``boundary Yang-Baxter equation'' (3.25)
and (3.33); therefore there is no ordering problem in (3.38). The
commutativity (3.39) is crucial for the interpretation of $\Psi$ in
(3.38) as the ``wave function'' of the boundary state.

The ``crossing equations'' (3.33), (3.35) and the expression (3.38) for
the boundary state are the main results of this Section. We feel that the
simple universal form (3.38) of the boundary state is not accidental.
However at the moment we do not have a satisfactory understanding of its
profound meaning and we can not offer anything better than the direct
derivation through (3.31).

$4'$.Now we turn to the ``boundary bootstrap conditions''. There are two
sorts of these: the bootstrap conditions describing the boundary
scattering of the ``bound-state'' particles and the conditions related to
possible existence of the ``boundary bound states''.

If the particle $A_c$ can be interpreted as the bound state of $A_a A_b$
(i.e. the pole at $\theta = iu_{a b}^c$, with $u_{a b}^c$ satisfying
(3.11)), the boundary S-matrix elements $R_{c}^{d}(\theta)$ can be
obtained by taking the appropriate residue in the bound-state pole of
the two-particle boundary S-matrix $R_{a b}^{a b}(\theta_1, \theta_2)$. This
way one gets the equation

$$f_{d}^{a b}R_{c}^{d}(\theta)= f_{c}^{b_1 a_1}R_{a_1}^{a_2}(\theta +
i\bar u_{a d}^{b})S_{b_1
a_2}^{b_2 a}(2\theta + i\bar u_{a d}^{b} - i \bar u_{b
d}^{a})R_{b_2}^{b}(\theta - i\bar u_{b d}^{a} ) \eqno(3.40)$$
($\bar u \equiv i\pi - u$) which has the diagrammatic representation
shown in Fig.13. If the particles $A_a$ and $A_b$ have equal masses
one can expect the appearence of a pole of
$R_{\bar a}^{b}(\theta)$ at $\theta = {i\pi \over 2}- {u_{a
b}^c \over 2} = \bar u_{a b}^c - {i\pi \over 2}$ due to the diagram in
Fig.14. The corresponding residue can be written as

$$K^{a b}(\theta) \simeq {i\over 2}\,{{f_{c}^{a b} g^{c}}\over
{\theta - iu_{a b}^c}}, \eqno(3.41)$$
where the ``three-particle couplings'' $f$ are the same as in (3.12) but
$g^c$ are new constants describing the ``couplings'' of the particles
$A_c$ to the boundary (Fig.14). More precisely, the nonzero value of $g^c$
indicates that the boundary state $\mid B \rangle$ contains a separate
contribution of the zero-momentum particle $A_c$, i.e.

$$\mid B \rangle = {\cal N}(1 + g^c A_c (0) + {1\over
2}\int_{-\infty}^{\infty}d\theta K^{a b}(\theta)A_a (-\theta)A_b
(\theta) + ... )\mid 0 \rangle. \eqno(3.42)$$
Let us stress again that in our previous analysis of the boundary state
we have ignored this possibility, i.e. strictly speaking validity of the
equation (3.38) above is limited to the case when $g^c = 0$ for all the
particles $A_c$ in the theory. It is easy to show that

$$[g^c A_c (0), K(\theta)] = 0, \eqno(3.43)$$
where $K(\theta)$ is the bilinear operator (3.39), and so in the general
case one can look for the ``boundary wave function'' in the form
$\Psi[g^c A_c (0), K(\theta)]$. The commutativity (3.43) follows from the
relation

$$g^{c'}K^{a' b}(\theta)S_{c' a'}^{c a}(\theta) = g^{c'}K^{a b'}(\theta)
S_{b' c'}^{c b}(\theta) \eqno(3.44)$$
(Fig.15) which is easily obtained if one considers the limit $\theta \to
{i\pi \over 2} - {1\over 2}u_{a_1 b_1}^{c}$ in (3.25) and takes into
account (3.13). It is also possible to show that if $g^{c_1}, g^{c_2}
\neq 0$ the amplitude $K^{c_1 c_2}(\theta)$ has the pole at $\theta = 0$
with the residue

$$K^{c_1 c_2}(\theta) \simeq -{i\over 2}\,{{g^{c_1} g^{c_2}}\over \theta};
\eqno(3.45)$$
This pole term is illustrated by the diagram in Fig.16.

Let us illustrate these bootstrap conditions with an example of the
so-called ``Lie-Yang field theory''. This field theory describes the
scaling limit of Ising Model with purely imaginary external field, the
critical point being the ``Lie-Yang edge singularity''(see [18]). It is
a massive
field theory which can be obtained by perturbing the $c= -{22\over 5}$
minimal CFT by its only nontrivial primary field $\varphi \equiv
\Phi_{(1,2)}$, which has a conformal dimension $\Delta = -{1\over 5}$.
Mussardo and Cardy[11] have shown that the ``bulk'' theory is integrable;
they have also found the corresponding factorizable ``bulk'' S-matrix.
The theory contains only one species of particles, $A$, and the ``bulk''
S-matrix is described by

$$A(\theta_1)A(\theta_2) = S(\theta_1 -
\theta_2)A(\theta_2)A(\theta_1) \eqno(3.46)$$
with

$$S(\theta) = {{\sinh\theta + i\sin{{2\pi}\over 3}}\over{\sinh\theta -
i\sin{{2\pi}\over 3}}}. \eqno(3.47)$$
The pole of $S(\theta)$ at $\theta = {{2\pi i}\over 3}$ ( which has
negative residue thus making non-unitarity of this theory manifest)
corresponds to the same particle $A$ appearing as the ``$AA$ bound state''.
Here we do not attempt to analyze the possible integrable boundary
conditions in this theory; we just assume that such ones do exist.
Then the factorizable boundary S-matrix is described by (3.46) and

$$A(\theta)B = R(\theta)A(-\theta)B, \eqno(3.48)$$
where the boundary scattering amplitude $R$ has to satisfy (3.26) and
(3.35) (the boundary Yang-Baxter equation (3.25) is satisfied
identically), i.e.

$$R(\theta)R(-\theta) = 1; \quad K(\theta) = S(2 \theta)K(-\theta); \quad
K(\theta) = R({{i\pi}\over 2} - \theta). \eqno(3.49)$$
As the particle $A$ appears as the ``bound state'', the equation (3.40)
has to be imposed as well,

$$R(\theta) = S(2\theta)R(\theta + {{i\pi}\over 3})R(\theta -
{{i\pi}\over 3}). \eqno(3.50)$$
There are two ``minimal'' solutions to (3.49), (3.50)\footnote{$^6$}{We
call ``minimal'' the solution which has ``minimal'' set of zeroes and
poles in the physical strip $0\leq Im\theta \leq \pi$, i.e. one can not
reduce the total number of zeroes and poles in this strip without
violating (3.49), (3.50).},

$$R_{(1)}(\theta) = -{{\sinh({\theta\over 2} +{{i\pi}\over
4})}\over{\sinh({\theta\over 2} - {{i\pi}\over 4})}}{{\sinh({\theta\over
2} +{{i\pi}\over 12})}\over{\sinh({\theta\over 2} - {{i\pi}\over
12})}}{{\sinh({\theta\over 2} - {{i\pi}\over
3})}\over{\sinh({\theta\over 2} + {{i\pi}\over 3})}},\eqno(3.51)$$

$$R_{(2)}(\theta) = {{\sinh({\theta\over 2} -{{i\pi}\over
4})}\over{\sinh({\theta\over 2} + {{i\pi}\over 4})}}{{\sinh({\theta\over
2} -{{5i\pi}\over 12})}\over{\sinh({\theta\over 2} + {{5i\pi}\over
12})}}{{\sinh({\theta\over 2} - {{i\pi}\over
3})}\over{\sinh({\theta\over 2} + {{i\pi}\over 3})}} \eqno(3.52)$$
(of course $R_{(1)}$ and $R_{(2)}$ are related through the ``CDD factor''
(3.27)). These two solutions have rather different physical properties.
$R_{(1)}(\theta)$ exhibits a pole at $\theta = {{i\pi}\over 6}$
associated with the diagram in Fig.14. This means that for this solution
the boundary state $\mid B \rangle_{(1)}$ contains the single-particle
contribution

$$\mid B \rangle_{(1)} = (1 + g_{(1)}A(0) + ...)\mid 0 \rangle
\eqno(3.53)$$
with non-zero amplitude $g_{(1)}$. Explicit calculation gives

$$g_{(1)} = 2i\sqrt{2\sqrt{3}-3}. \eqno(3.54)$$
Accordingly, the amplitude $R_{(1)}(\theta)$ has a pole at $\theta =
{{i\pi}\over 2}$ with the residue

$$R_{(1)}(\theta) \simeq {i{g_{(1)}^2}\over {2\theta} - {i\pi}}.
\eqno(3.55)$$
The solution $R_{(2)}$ does not have any poles in the physical strip,
i.e. $g_{(2)} = 0$.

In the general case the poles (3.41) and (3.45) do not exhaust all possible
singularities the amplitudes $R_{a}^{b}(\theta)$ may have in the
``physical strip'' $0 \leq Im\theta \leq \pi$. Even with given boundary
action (2.8)((2.9)) the boundary can exist in several stable states $\mid
\alpha \rangle_B; \quad \alpha = 0, 1, ..., n_B-1$ (let us stress that
all these states belong to ${\cal H}_B$). These states are eigenstates
of the operators (2.16)

$$H_s \mid \alpha \rangle_B = h_{\alpha}^{(s)} \mid \alpha \rangle_B,
\eqno(3.56)$$
with some eigenvalues $h_{\alpha}^{(s)}$; in particular $e_{\alpha}=
h_{\alpha}^{(1)}$ is the energy of the state $\mid \alpha \rangle_B$.
We assume that $e_{0}$ is the smallest of $e_{\alpha}$, i.e. as before
the state $\mid 0 \rangle_B$ is the ground state of
$H_B$\footnote{$^7$}{Here we assume that the ground state is not
degenerate.}. If $e_{\alpha}-e_0 <
\min m_a$ the state $\mid \alpha \rangle_B$ is stable just because its decay
is forbidden energetically; higher IM $H_s$ could ensure stability of
$\mid \alpha \rangle_B$ even if $e_{\alpha}-e_0 > \min m_a$. In any
case, the states $\mid \alpha \rangle_B$ must show up as the virtual
states in the boundary scattering processes. The amplitudes
$R_{a}^{b}(\theta)$ can exhibit the poles

$$R_{a}^{b}(\theta) \simeq {i\over 2}\,{{g_{a 0}^{\alpha}g_{\alpha}^{b
0}}\over {\theta - iv_{0 a}^{\alpha}}}, \eqno(3.57)$$
where $g_{a 0}^{\alpha}$ are ``boundary-particle couplings'' and
$v_{0 a}^{\alpha}$ satisfy

$$e_0 + m_a \cos(v_{0 a}^{\alpha}) = e_{\alpha}, \eqno(3.58)$$
see Fig.17.
Thus the states $\mid \alpha \rangle_B$ with $e_{\alpha} > e_0$ can be
interpreted as the ``boundary bound states'', $e_{\alpha}- e_0$ being the
binding energy.

In this situation it is natural to generalize (3.22) to

$$A_a (\theta)B_{\alpha} = R_{a \alpha}^{b \beta}(\theta)A_b (-\theta)
B_{\beta}, \eqno(3.59)$$
where $B_{\alpha}$ are formal ``boundary creating operators'' analogous
to (3.21) and $R_{a \alpha}^{b \beta}(\theta)$ are  amplitudes of the
scattering process involving the change of the state of boundary, as
shown in Fig.18. Again, the amplitude $R_{a \alpha}^{b \beta}(\theta)$
does not vanish only if $m_a = m_b$ and $e_{\alpha} = e_{\beta}$, as it
follows from (3.56). The equations (3.25), (3.26), (3.33), (3.35),
(3.40) above can be generalized in an obvious way to include these
amplitudes. Here we quote only the ``boundary bound state bootstrap
equation''

$$g_{a \alpha}^{\gamma}R_{b \gamma}^{b' \beta}(\theta) =
g_{a_2 \gamma}^{\beta}S_{b a}^{b_1 a_1}(\theta - iv_{a \alpha}^{\beta})
R_{b_1 \alpha}^{b_2 \gamma}(\theta)S_{a_1 b_2}^{a_2 b'}
(\theta + iv_{a \alpha}^{\beta}) \eqno(3.60)$$
shown in Fig.19; here the parameter $v_{a \alpha}^{\beta}$ satisfies the
equation

$$e_{\alpha} + m_a \cos (v_{a \alpha}^{\beta}) = e_{\beta}, \eqno(3.61)$$
analogous to (3.38), and $g_{a \alpha}^{\beta}$ are the ``particle-boundary
coupling constants'' (Fig.20) entering the residues

$$R_{a \alpha}^{b \beta}(\theta) \simeq {i\over 2}\,
{{g_{a \alpha}^{\gamma}
g_{\gamma}^{b \beta}}\over {\theta - iv_{a \alpha}^{\gamma}}}.
\eqno(3.62)$$
Note that (3.44) and (3.45) can be considered as particular cases
$\alpha = \beta = \gamma = 0$ of (3.60) and (3.62), resp., if we think
of the ground state $\mid 0 \rangle_B$ as the ``bound state'' of some
particle $A_c$ (with $g^c \neq 0$) and itself.

We have implicitely assumed in the above discussion that the ground
state $\mid 0 \rangle_B$ is non-degenerate. This assumption is taken for
the sake of simplicity only; it is not nesessary either from a physical
point of view or for mathematical consistency. It is
straightforward to incorporate the possibility of degenerate ground states
into the above picture. In the next two Sections, where we consider
examples of the factorizable boundary scattering theory, we will encounter
this interesting possibility.

\vskip 0.4in

4.ISING MODEL.

As is known, the scaling limit of the Ising Model with zero external field
is described by the free Majorana fermion field theory

$${\cal A} = \int dy dx \quad {a}_{FF}(\psi, \bar \psi), \eqno(4.1)$$
where

$$a_{FF}(\psi, \bar \psi) = \psi \partial_{\bar z} \psi -
{\bar \psi} \partial_{z}{\bar \psi} + m \psi {\bar \psi}. \eqno(4.2)$$
Here $(z, \bar z)$ are complex coordinates and $m \simeq T_c - T$.
The field theories corresponding to the high-temperature ($T-T_c \to 0^{+}$)
and the low temperature ($T-T_c \to 0^{-}$) phases of the Ising Model are
equivalent (they are related through the duality transformation $\psi \to
\psi; \quad \bar \psi \to - \bar \psi$, which changes the sign of $m$ in
(4.1)). Here we will assume that $m > 0$ and interpret this field theory
as the low-temperature phase. In this phase there are two degenerate
ground states, $\mid 0, \pm \rangle$, so that the corresponding
expectation values of the spin field $\sigma(x)$ are $\langle \sigma(x)
\rangle_{\pm} = \pm \bar \sigma$, where $\bar \sigma \sim m^{1\over 8}$
is the spontaneous magnetization.

The bulk theory (4.1) contains one sort of particles - the free fermion
$A$ with the mass $m$. Intuitively, this particle can be understood as a
``kink'' (or ``domain wall'') separating domains of opposite
magnetization. The corresponding particle creation operator $A^{\dagger}
(\theta)$ (we denote it here $A^{\dagger}$ to comply with the conventional
notations) can be defined through the decomposition

$$\psi(x, t) = \int_{-\infty}^{\infty}d\theta [\omega e^{\theta \over 2}
A(\theta) e^{imx\sinh\theta + imt\cosh\theta}+\bar \omega e^{\theta \over 2}
A^{\dagger}(\theta) e^{-imx\sinh\theta - imt\cosh\theta}];$$

$$\bar \psi(x, t) = \int_{-\infty}^{\infty}d\theta [\bar \omega
e^{-{\theta \over 2}}
A(\theta) e^{imx\sinh\theta + imt\cosh\theta}+\omega e^{-{\theta \over 2}}
A^{\dagger}(\theta) e^{-imx\sinh\theta - imt\cosh\theta}], \eqno(4.3)$$
where $t= iy$ and $\omega = \exp({{i\pi}\over 4}); \quad \bar \omega =
\exp(-{{i\pi}\over 4})$. The operators $A(\theta), A^{\dagger}(\theta)$
satisfy canonical anticommutation relations $\lbrace A(\theta),
A^{\dagger}(\theta')\rbrace = \delta(\theta - \theta');\,\lbrace
A(\theta), A(\theta')\rbrace = \lbrace A^{\dagger}(\theta),
A^{\dagger}(\theta')\rbrace$ $ = 0$. The last of these relations,

$$A^{\dagger}(\theta)A^{\dagger}(\theta')= - A^{\dagger}(\theta')
A^{\dagger}(\theta), \eqno(4.4)$$
has the same meaning as (3.3) so that the free-fermion two particle
S-matrix is [19]

$$S = -1. \eqno(4.5)$$

Let us consider now this field theory in the half-plane $x < 0$, with
the boundary at $x = 0$. Assuming that the boundary conditions are
chosen to be integrable, we can define the boundary scattering amplitude
$R(\theta)$,

$$A^{\dagger}(\theta) B = R(\theta) A^{\dagger}(-\theta) B. \eqno(4.6)$$
As is discussed in Sect.3, this amplitude has to satisfy (3.26), (3.33),
(3.35), i.e.

$$R(\theta)R(-\theta) = 1; \qquad K(\theta) = - K(-\theta);$$
$$K(\theta) = R({{i\pi}\over 2} -\theta). \eqno(4.7)$$
We consider first two simplest boundary conditions - the ``free'' and
the ``fixed'' ones\footnote{$^8$}{In the conformal field theory of the Ising
model
(which describes the case $m=0$), these are just the two possible conformal
boundary conditions; they are analyzed in [7,8].}.

a). ``Fixed'' boundary condition. In the microscopic theory this boundary
condition corresponds to fixing the boundary spins to be, say, $+1$.
Obviously, this boundary condition removes the ground state degeneracy.
In terms of the fermion fields $\psi, \bar \psi$ the ``fixed'' boundary
condition can be written as

$$(\psi + \bar \psi)_{x=0} = 0. \eqno(4.8)$$
In presence of the boundary, the fields $\psi, \bar \psi$ still enjoy the
decomposition (4.3), although the operators $A, A^{\dagger}$ (now acting
in ${\cal H}_B$) are not all independant but satisfy the relations

$$(\omega e^{\theta \over 2} + \bar \omega e^{-{\theta \over
2}})A(\theta) = -(\bar \omega e^{\theta \over 2} + \omega  e^{-{\theta
\over 2}})A(-\theta);$$
$$(\bar \omega e^{\theta \over 2} + \omega  e^{-{\theta
\over 2}})A^{\dagger}(\theta) = - (\omega e^{\theta \over 2} + \bar
\omega e^{-{\theta \over 2}})A^{\dagger}(-\theta) \eqno(4.9)$$
which follow from (4.8). From the last of (4.9) we find

$$R_{fixed}(\theta) = i\tanh({{i\pi}\over 4} - {\theta \over 2}).
\eqno(4.10)$$
Alternatively, one could use another hamiltonian picture, with $\tau =
-ix$ interpreted as the time. In this picture the fields $\chi = \omega
\psi, \bar \chi = \bar \omega \bar \psi$ enjoy the same decomposition
(4.3) with the substitution $ x \to y, t \to \tau$ and operators
$A(\theta), A^{\dagger}(\theta)$ acting in the Hilbert space $\cal H$
of the
bulk theory. The boundary condition (4.6) appears as the initial condition
$(\chi + i\bar \chi)_{\tau = 0} = 0$ which is understood as the equation

$$(\chi + i\bar \chi)_{\tau = 0}\mid B_{fixed} \rangle = 0 \eqno(4.11)$$
for the corresponding boundary state $\mid B_{fixed} \rangle$.
In terms of $A, A^{\dagger}$ this equation reads

$$[\cosh({\theta/ 2})A(\theta) + i\sinh({\theta/
2})A^{\dagger}(-\theta)]\mid B_{fixed} \rangle = 0, \eqno(4.12)$$
and hence the boundary state can be written as

$$\mid B_{fixed} \rangle = \exp \lbrace {1\over
2}\int_{-\infty}^{\infty} d\theta K_{fixed}(\theta)A^{\dagger}(-\theta)
A^{\dagger}(\theta)\rbrace \mid 0 \rangle \eqno(4.13)$$
with

$$K_{fixed}(\theta) = i\tanh{\theta \over 2} \eqno(4.14)$$
and $\mid 0 \rangle = \mid 0,+ \rangle$. Although (4.13) with $\mid 0
\rangle = \mid 0,- \rangle$ solves (4.12) as well, it is easy to see
that in the infinite system this state does not contribute to $\mid
B_{fixed} \rangle$ (in a large system, finite in the $y$-direction,
$-L/2<y<L/2$, its contribution is suppressed as $\exp(-mL)$). Note that
(4.10), (4.14) satisfy (4.7).

b). ``Free'' boundary condition. In the microscopic theory one imposes no
restrictions on the boundary spins. Correspondingly, the ground state is
still two-fold degenerate. Again, in terms of the fermions $\psi, \bar \psi$
this boundary condition has a very simple form

$$(\psi - \bar \psi)_{x=0} = 0. \eqno(4.15)$$
The same computation as in the previous case gives

$$R_{free}(\theta) = -i\coth({{i\pi}\over 4} - {\theta \over 2}).
\eqno(4.16)$$
Note that the corresponding boundary state amplitude

$$K_{free}(\theta) = -i\coth{\theta \over 2} \eqno(4.17)$$
exhibits a pole at $\theta = 0$ (this pole is related to the existence of
the zero-energy mode $\psi = \bar \psi \sim \exp{mx}$ which satisfies (4.15)),
which indicates that the boundary state $\mid B_{free} \rangle$
contains the contribution of a zero-momentum one-particle state,

$$\mid B_{free} \rangle = (1 + A^{\dagger}(0) + ...)\mid 0 \rangle.
\eqno(4.18)$$
Of course, this feature is easily understood in
physical terms. Semi-infinite Ising Model at $T , T_c$ with free
boundary condition admits a particular equilibrum state, characterized by
the asymptotic conditions $\langle \sigma(x,y) \rangle \to +\bar \sigma$ as
$y \to +\infty$ and  $\langle \sigma(x,y) \rangle \to -\bar \sigma$ as
$y \to -\infty$. This state contains an infinitely long (fluctuating)
``domain wall'' attached to the boundary, which separates two domains
of opposite magnetization, as shown in Fig.21. This ``domain wall''
configuration is interpreted as a zero-momentum particle emitted by the
boundary state. It is not difficult to check that the boundary state

$$\mid B_{free} \rangle = (1 + A^{\dagger}(0))\exp \lbrace {1 \over 2}
\int_{-\infty}^{\infty}
K_{free}(\theta)A^{\dagger}(-\theta)A^{\dagger}(\theta)\rbrace \mid 0
\rangle \eqno(4.19)$$
satisfies the boundary state equation

$$\int d\theta f(\theta)[\sinh(\theta/2)A(\theta) -
i\cosh(\theta/2)A^{\dagger}(-\theta)]\mid B_{free} \rangle = 0
\eqno(4.20)$$
($f(\theta)$ is an arbitrary smooth function) which follows from the
``initial condition'' $(\chi -i\bar \chi)_{\tau=0} =0$.

The simple boundary conditions above can be obtained as the two limiting
cases of the more  general integrable boundary condition which we describe
below.

c). ``Boundary magnetic field''. This boundary condition is obtained by
introducing a nonzero external field $h$ (``boundary magmetic field'')
which couples only to the boundary spins. Obviously,  $h=0$ correspond
to the ``free'' case above while in the limit $h \to \infty$ one
recovers the ``fixed'' boundary condition. The boundary magnetic field
can be considered as the perturbation of the ``free'' boundary condition,

$${\cal A}_{h} = {\cal A}_{free} + h \int_{-\infty}^{\infty}dy \sigma_B
(y), \eqno(4.21)$$
where $\sigma_B (y)$ is the ``boundary spin operator''
[7,8]\footnote{$^9$}{This perturbation generates a ``flow'' from the free
boundary condition down to the fixed one. In the case of conformal ``bulk''
theory ($m=0$) this flow was studied in [20].}. The action ${\cal
A}_{free}$ is

$${\cal A}_{free} = \int_{-\infty}^{\infty}dy \int_{-\infty}^{0}dx
a_{FF}(\psi, \bar \psi) + {1/2}\int_{-\infty}^{\infty}dy [(\psi \bar
\psi)_{x=0} + a\dot a]. \eqno(4.22)$$
Here $\dot a ={d\over dy}a$, and $a(y)$ is an additional (fermionic)
boundary degree of freedom which is introduced to describe the
ground-state degeneracy. The quantum operator $a$ associated with the
boundary field $a(y)$ anticommutes with $\psi, \bar \psi$ and satisfies

$$a^2 = 1, \eqno(4.23)$$
so that for the free boundary condition one has

$$a\mid 0, \pm \rangle_B = \mid 0, \mp \rangle_B . \eqno(4.24)$$
The equation (4.15) follows directly from the action (4.22). The boundary
spin operator $\sigma_B (y)$ is analyzed (along with other boundary
operators) in [7,8]. It is identified with the degenerate primary boundary
field $\psi_{(1,3)}$ of dimension $\Delta = 1/2$. In the Lagrangian picture
described above it can be written as

$$\sigma_B (y) = {1\over 2}(\psi + \bar \psi)_{x=0}(y)a(y), \eqno(4.25)$$
so that the dimension of $h$ in (4.21) is $[mass]^{1/2}$.

 One obtains, from (4.21) and (4.25), the boundary condition for the fermi
fields $\psi, \bar \psi$,

$$i{d\over dy}(\psi - \bar \psi)_{x=0} = {{h^2}\over 2}(\psi + \bar
\psi)_{x=0}. \eqno(4.26)$$
With this, it is straightforward to compute the boundary scattering
amplitude for (4.21),

$$R_{h}(\theta)=i\tanh({{i\pi}/ 4} - {\theta/ 2}){{\kappa -
i\sinh\theta}\over {\kappa + i\sinh\theta}}, \eqno(4.27)$$
where

$$\kappa = 1 - {{h^2}\over 2m}. \eqno(4.28)$$
Note that for $h > 0$ the amplitude $R_{h}(\theta)$ does not
have a pole at $\theta = {{i\pi}/2}$ and hence there is no one-particle
(and any odd-number particle) contributions to the corresponding
boundary state $\mid B_h \rangle$; this state has therefore the general
form (3.38) with $\mid 0 \rangle = \mid 0,\pm \rangle$, for $h=\pm |h|$,
resp.,  and with

$$K_{h}(\theta) = i\tanh(\theta /2){{\kappa + \cosh\theta}\over {\kappa
- \cosh\theta}}. \eqno(4.29)$$
Again, this is not very surprising. The nonzero boundary magnetic field
removes the ground-state degeneracy; for $h$ sufficiently small, the
two-fold degenerate ground state $\mid 0, \pm \rangle_{B}$ of the ``free''
boundary splits into two non-degenerate states $\mid 0 \rangle_{B}$ and
$ \mid 1 \rangle_B$, where $\mid 1 \rangle_B$ can be interpreted as
the boundary bound state. For $0 < h^2 < 2m$ we can parametrize (4.28) as

$$\kappa = \cos v. \eqno(4.30)$$
with real $v$, $0 < v <{\pi/2}$. In this domain the amplitude
(4.27) exhibits a pole in the physical strip at $\theta = i({\pi/2}-v)$
which is associated with the state $\mid 1 \rangle_B$. Physical
interpretation of this boundary bound state is simple. For $h\neq 0$ the
equilibrium ``ground state'' which minimizes the free energy features
an asymptotic behaviour $\langle \sigma (x,y) \rangle_{0} \to
+sign(h) \bar\sigma$ as $x \to -\infty$. However, for $h$ sufficiently small,
there exists another stable equilibrum state with $\langle \sigma (x,y)
\rangle_{1} \to -sign(h) \bar \sigma$ as $x \to -\infty$.
Although its free energy is higher by a positive boundary term, it
is indeed stable as in the infinite system there is no finite kinetics of
decay (in a system finite in the $y$ direction the decay
probability is suppressed as $\exp(-mL\cos v)$). Clearly, the energies
$e_0$ and $e_1$ of the states $\mid 0 \rangle_B$ and $\mid 1 \rangle_B$
have the meaning of specific (per unit boundary length) boundary free
energies of these two equilibrum states. From (4.27) we have

$$e_1 - e_0 = m\sin v. \eqno(4.31)$$
It is also clear that the expectation values $\langle ... \rangle_{1}$
can be obtained by analytic continuation of $\langle ... \rangle_{0}$
from $h$ to $-h$. So, in this very domain $|h|<\sqrt{2m}$ ``boundary
histeresis''[21] can be observed. It is possible to show that these
expectation values can be computed as the matrix elements

$$\langle ... \rangle_{1} = {{_{B}\langle 1 \mid ...\mid 1
\rangle_{B}}\over{_{B}\langle 1 \mid 1 \rangle_{B}}} =
{{\langle 0' \mid ...\mid B_{h}' \rangle}\over {\langle 0' \mid B_{h}'
\rangle}}, \eqno(4.32)$$
where the last expression contains the ``excited boundary state''

$$\mid B_{h}' \rangle = exp({1\over 2}\int_{\cal C} d\theta K_h
(\theta)A^{\dagger}(-\theta)A^{\dagger}(\theta))\mid 0'\rangle
\eqno(4.33)$$
with the integration taken along contour $\cal C$ shown in Fig.2, and $\mid
0'\rangle = \mid 0, \mp\rangle$ for $h= \pm |h|$.

As $h$ approaches the ``critical'' value $h_c = \sqrt{2m}$ (i.e. $v$
approaches $\pi/2$), the ``boundary bound state'' becomes weakly bound;
$e_1 - e_0 - m = -2m\sin^2 (\pi/4 -v/2) \to 0$, and its effective size
$\xi = (e_1 - e_0 - m)^{-1}$
diverges as $(h_c -h)^{-2}$. Correspondingly, the equilibrium state
$\langle ... \rangle_{1}$ developes large boundary fluctuations which
propagate deep into the bulk,

$$\langle \sigma(x,y) \rangle_{1} + \bar \sigma \simeq \exp(x/\xi) \quad
as \quad x \to -\infty. \eqno(4.34)$$
At $h=h_c$ the boundary condition (4.26) reduces to

$${\partial\over \partial x}(\psi + \bar \psi)_{x=0} = 0 \eqno(4.35)$$
and the boundary scattering amplitude can be written as

$$R_{h=h_c}(\theta) = -i\tanh({{i\pi}/4}-\theta/2) \eqno(4.36)$$
which differs only by a sign from (4.10). This difference is  very
significant though; (4.36) admits the state $A^{\dagger}(0)\mid 0 \rangle_B$
- remnant of $\mid 1 \rangle_B$.

For $2m < h^2 \leq 4m$ we can still  parametrize $\kappa$ as in (4.30),
 with real $v$. However, now ${\pi/2}\leq v \leq \pi$. The pole of
$R_h$ at $\theta = i({\pi/2}- v)$ leaves the physical strip; there is
no ``boundary bound state'' in this domain, and the equilibrum state is
unique. At $h=2\sqrt m$ the expression (4.27) simplifies as

$$R_{h=2\sqrt m}(\theta) = -i\tanh^{3}({{i\pi}/4}-\theta/2),, \eqno(4.37)$$
i.e. the boundary-state amplitude $K_{h=2\sqrt m}$ possesses a third-order
zero at $\theta = 0$. This seems to affect the $x \to -\infty$
asymptotic of $\langle \sigma(x,y) \rangle$, but at the moment we do not
have a clear physical interpretation of this phenomenon.

Finally, at $h^2 > 4m$ one can write $\kappa = -\cosh \theta_0$ with
$\theta_0$ real and positive. In this regime the amplitude $R_h$ shows
two ``resonance'' poles at $\theta = -{i\pi/2}\pm \theta_0$. As $h$
grows these poles depart to infinity and ultimately $R_{\infty}$ reduces
to (4.10).

Clearly, the behavior described above agrees with the expected
``flow'' from the
``free'' boundary condition down to the ``fixed'' one[20]. This example is
particularly simple because the described boundary field theory is not
only integrable but free - both the bulk and the boundary equations of motion
for $\psi, \bar \psi$ are linear. It is easy to show that all solutions
of (4.7) correspond to free-field boundary conditions, generalizing
(4.26) in an obvious way.

\vskip 0.4in

5.BOUNDARY SINE-GORDON MODEL

In this Section we consider the integrable boundary theory for the sine-Gordon
model. The bulk sine-Gordon field theory is described by the euclidean
action (2.1) with

$$a(\varphi, \partial_{\mu}\varphi ) =
{1\over 2}(\partial_{\mu}\varphi)^{2} - {{m^2}\over {\beta^2}}\cos(\beta
\varphi) \eqno(5.1)$$
where $\varphi(x,y)$ is a scalar field and $\beta$ is a dimensionless
coupling constant. This model is well known to be integrable both as
classical and quantum field theories (see e.g.[22]. In the quantum theory, the
discrete
symmetry $\varphi \to \varphi + {2\pi\over \beta}N, N\in {\bf Z}$ is
spontaneously broken at $\beta^2 < 8\pi$[23]; in this domain the theory is
massive and its particle spectrum consists of a soliton-antisoliton pair
$(A, \bar A)$ (with equal masses) and a number (which depends on $\beta$)
of neutral particles (``quantum breathers'') $B_n$, $n = 1, 2, ... <
\lambda$, where

$$\lambda = {{8\pi}\over {\beta^2}}-1. \eqno(5.2)$$
The soliton (antisoliton) carries a positive (negative) unit of ``topological
charge''

$$q = {\beta \over{2\pi}}\int_{-\infty}^{+\infty}dx
{\partial\over{\partial x}}\varphi (x,y) = {\beta
\over{2\pi}}[\varphi(+\infty, y)-\varphi(-\infty, y)]. \eqno(5.3)$$
The charge conjugation ${\bf C}: A \leftrightarrow \bar A$ is related to
$\varphi \leftrightarrow -\varphi$ symmetry of (5.1). The particles
$B_n$ are neutral (they are interpreted as the soliton-antisoliton bound
states), $B_n$ with even (odd) $n$ being $\bf C$ -even ($\bf C$ -odd).
The masses of $B_n$ are

$$m_n = 2M_{s}\sin({{n\pi}\over{2\lambda}}); \qquad n = 1, 2, ... < \lambda,
\eqno(5.4)$$
where $M_s$ is the soliton mass.

Factorizable scattering of solitons is described by the commutation
relations
$$A(\theta)A(\theta') = a(\theta-\theta')A(\theta')A(\theta), \qquad
\bar A(\theta)\bar A(\theta') =
a(\theta-\theta')\bar A(\theta')\bar A(\theta),$$
$$A(\theta)\bar A(\theta') = b(\theta-\theta')\bar A(\theta')A(\theta) +
c(\theta - \theta')A(\theta')\bar A(\theta), \eqno(5.5)$$
where $A(\theta)$ and $\bar A(\theta)$ are soliton and antisoliton
creation operators and the two-particle scattering amplitudes $a, b, c$
are given by

$$a(\theta) = \sin(\lambda (\pi - u))\rho (u),$$
$$b(\theta) = \sin(\lambda u)\rho (u), \eqno(5.6)$$
$$c(\theta) = \sin(\lambda \pi)\rho (u),$$
where $u = -i\theta$ and

$$\rho(u) = -{1\over\pi}\Gamma({\lambda})\Gamma(1-{{\lambda
u}/\pi})\Gamma(1-\lambda+{{\lambda
u}/\pi})\prod_{l=1}^{\infty}{{F_{l}(u)F_{l}(\pi - u)}\over
{F_{l}(0)F_{l}(\pi)}};$$
$$F_{l}(u) = {{\Gamma(2l\lambda -{\lambda u/\pi})\Gamma(1+2l\lambda
-{\lambda
u/\pi})}\over{\Gamma((2l+1)\lambda-{\lambda
u/\pi})\Gamma(1+(2l-1)\lambda-{\lambda u/\pi})}}. \eqno(5.7)$$
The amplitudes of $A B_n$ and $B_n B_m$ scatterings can be obtained from
these with the use of (3.13), they can be found in [4].

Let us consider now this field theory in presence of the boundary at
$x=0$, described by the action (2.8). The first question which arises
here is how to choose the ``boundary action'' $b(\varphi_B,
{d/dy}\varphi_B)$ in (2.8) in order to preserve integrability. General
classification of integrable boundary actions seems to be an interesting
open problem. Here we just claim that the boundary sine-Gordon theory (2.8),
(5.1) with

$$b(\varphi) = -M\cos({\beta\over 2}(\varphi - \varphi_0)), \eqno(5.8)$$
where $M$ and $\varphi_0$ are free parameters, is integrable. This claim is
supported by an explicit computation of the first nontrivial integral of
motion in the Appendix, where we restrict attention to the classical case; we
believe this computation can be extended to the quantum theory along the
lines explained in Sect.2. We want to describe the factorizable boundary
scattering theory associated with (5.8).

The soliton (antisoliton) boundary scattering amplitudes can be encoded
in the commutation relations (see Section 2)

$$A(\theta)B = P_{+}(\theta)A(\theta)B + Q_{+}(\theta){\bar
A}(\theta)B;$$
$${\bar A}(\theta)B = P_{-}(\theta){\bar A}(\theta)B + Q_{-}(\theta)
A(\theta)B. \eqno(5.9)$$
Here $P_{+}$, $Q_{+}$ ($P_{-}$, $Q_{-}$) are the amplitudes of
soliton (antisoliton) one-particle boundary scattering processes shown
in Fig.23. Some comments are in order. Exept for the case $M=\infty$ (see
below), the boundary value $\varphi(x=0,y)$ is not fixed in the boundary
field theory (5.8)and hence the topological charge

$$q = {\beta \over{2\pi}}\int_{-\infty}^{0}dx
{\partial\over{\partial x}}\varphi (x,y) \eqno(5.10)$$
is not conserved. Therefore the boundary scatterings are allowed to
violate this charge conservation by an even number. In particular, an incoming
soliton can go away as an antisoliton and vice versa. The
amplitudes $Q_{\pm}$ are introduced to describe these processes.
The excceptional case is $M=\infty$ (``fixed'' boundary condition);
in this case we must have $Q_{\pm} = 0$. Also, at $\varphi_0 \neq
0(mod {{2\pi}/\beta})$ (and
$M \neq 0$) the boundary interaction violates charge-conjugation
symmetry; that is why in the general case the amplitudes  $P_{+}$, $Q_{+}$
and $P_{-}$, $Q_{-}$ are expected to be different.

With (5.9), the boundary Yang-Baxter equation (3.25) leads to six
distinct functional equations,

$$Q_{+}(\theta')c(\theta + \theta')Q_{-}(\theta)a(\theta - \theta') =
Q_{-}(\theta')c(\theta + \theta')Q_{+}(\theta)a(\theta -\theta'),
\eqno(5.11a)$$
$$P_{-}(\theta')b(\theta +\theta')P_{+}(\theta)c(\theta -\theta')+
P_{+}(\theta')c(\theta +\theta')P_{+}(\theta)b(\theta -\theta') =$$
$$P_{-}(\theta')c(\theta +\theta')P_{-}(\theta)b(\theta -\theta')+
P_{+}(\theta')b(\theta +\theta')P_{-}(\theta)c(\theta -\theta'),
\eqno(5.11b)$$
$$P_{+}(\theta')a(\theta + \theta')Q_{+}(\theta)c(\theta - \theta') +
Q_{+}(\theta')b(\theta + \theta')P_{+}(\theta)b(\theta - \theta') +
Q_{+}(\theta')c(\theta + \theta')P_{-}(\theta)c(\theta - \theta') =$$
$$Q_{+}(\theta')a(\theta +\theta')P_{+}(\theta)a(\theta -\theta')+
P_{-}(\theta')c(\theta +\theta')Q_{+}(\theta)a(\theta -\theta')
\eqno(5.11c)$$
$$P_{+}(\theta')a(\theta +\theta')Q_{+}(\theta)b(\theta -\theta')+
Q_{+}(\theta')b(\theta +\theta')P_{+}(\theta)c(\theta -\theta')+
Q_{+}(\theta')c(\theta +\theta')P_{-}(\theta)b(\theta -\theta') =$$
$$P_{+}(\theta')b(\theta +\theta')Q_{+}(\theta)a(\theta -\theta'),
\eqno(5.11d)$$
(the other two are obtained from (5.11c), (5.11d) by $+\leftrightarrow -$
substitution). The solution to these equations is found in a
recent paper[24]\footnote{$^2$}{We learned about [24] after we had
found this solution independently.}; it reads

$$P_{+}(\theta) = \cos(\xi +\lambda u)R(u);$$
$$P_{-}(\theta) = \cos(\xi -\lambda u)R(u);$$
$$Q_{+}(\theta) = {k_{+}\over 2}\sin(2\lambda u)R(u);$$
$$Q_{-}(\theta) = {k_{-}\over 2}\sin(2\lambda u)R(u), \eqno(5.12)$$
where again $u= -i\theta$; $\xi, k_{\pm}$ are free
parameters and $R(u)$ is an arbitrary function\footnote{$^3$}{This is
the general solution for generic $\lambda$. For integer $\lambda$ there are
additional solutions.}.

The free parameters in (5.12) have to be related to the parameters of
the boundary action. The solution (5.12) seems to contain one
parameter more than the action (5.8). It is easy to see however that one
of the two parameters $k_{\pm}$ in (5.12) can be removed by an appropriate
gauge transformation

$$A(\theta) \to e^{i\alpha} A(\theta),  {\bar A}(\theta) \to e^{-i\alpha}
{\bar A}(\theta) \eqno(5.13)$$
which leaves all the charge-conserving amplitudes unchanged and
transforms the amplitudes $Q_{\pm}$ as

$$Q_{+}(\theta) \to e^{-2i\alpha}Q_{+}(\theta); \quad Q_{-}(\theta) \to
e^{+2i\alpha}Q_{-}(\theta). \eqno(5.14)$$
Considered from the Lagrangian point of view, this transformation amounts to
adding a
total derivative term to the boundary action density (5.8),

$$b(\varphi) \to b(\varphi) + {{2\pi \alpha}\over
\beta}{d\over{dy}}\varphi. \eqno(5.15)$$
We use this freedom to set $k_{+} = k_{-} = k$. This leaves us with two
parameters, $\xi$ and $k$, which correspond to the two-parameters $M$
and $\varphi_0$ in (5.8).

The function $R(u)$ in (5.12) can be determined with the use of the``boundary
unitarity'' and the ``boundary cross-unitarity'' equations (3.26) and
(3.35). These equations reduce to

$$R(u)R(-u) = [\cos^2(\xi) - \sin^2(\lambda u) - k^2\sin^2(\lambda
u)\cos^2(\lambda u)]^{-1}; \eqno(5.16)$$
$$K(u) = \sin\lambda(\pi + 2u) \rho(2u) K(-u);\quad K(u) = R({\pi/
2}-u). \eqno(5.17)$$
It is not difficult to solve these equations. In fact, it is convenient
to solve (5.16) and (5.17) ``separately''. Let us factorize the function
$R(u)$ as

$$R(u)=R_{0}(u)R_{1}(u); \quad
K(u)=K_{0}(u)K_{1}(u), \eqno(5.18)$$
where $K_{0}(u)=R_{0}({\pi/2}-u)$, $K_{1}(u)=R_{1}({\pi/2}-u)$ so that these
factors satisfy

$$R_{0}(u)R_{0}(-u)=1;\quad K_{0}(u)= \sin\lambda(\pi + 2u) \rho(2u)K_{0}(-u)
\eqno(5.19)$$
and

$$R_{1}(u)R_{1}(-u)= [\cos^2(\xi) - (1+k^2)\sin^2(\lambda u) +
k^2\sin^4(\lambda u)]^{-1}; \quad K_{1}(u) = K_{1}(-u). \eqno(5.20)$$
The equations (5.19) do not contain the boundary parameters $\xi$ and
$k$; the solution can be written as

$$R_{0}(u)=F_{0}(u)/F_{0}(-u);$$
$$F_{0}(u)={{\Gamma(1-{{2\lambda u}/\pi})}\over {\Gamma(\lambda
-{{2\lambda u}/\pi})}}\times$$
$$\prod_{k=1}^{\infty}{{\Gamma(4\lambda k -
{{2\lambda u}/\pi})\Gamma(1 + 4\lambda k -{{2\lambda u}/\pi})
\Gamma(\lambda(4k+1))\Gamma(1+\lambda(4k-1))}
\over{\Gamma(\lambda(4k+1) -{{2\lambda u}/\pi})\Gamma(1 + \lambda
(4k-1) -{{2\lambda u}/\pi})\Gamma(1+4\lambda k)\Gamma(4\lambda k)}}
\eqno(5.21)$$
The factor $R_{0}$ contains the poles in the ``physical
strip'' $0 < u < {\pi/2}$ located at $u = u_n = n{\pi/{2\lambda}};
n = 1, 2, ... < \lambda$. Appearence of these poles is very well
expected. In the generic case, the boundary state $\mid B \rangle$
associated with the boundary condition (5.8) is expected to contain the
contributions of the zero-momentum particles $B_n$, which leads to the poles
(3.41). So, these poles of $R_{0}$ correspond to the diagrams in Fig.24.
Note that these poles should not appear in the amplitudes $Q_{\pm}$ and
the factor $\sin(2\lambda u)$ in (5.12) takes care of this.

The equation (5.20) contains all the information about the boundary
condition (i.e. the parameters $\xi$ and $k$). Its solution can be
written as

$$R_{1}(u) = {1\over {\cos\xi}}\sigma(\eta,u)\sigma(i\vartheta,u) \eqno(5.22)$$
where

$$\sigma(x,u) ={{\Pi(x, {\pi/2}-u)\Pi(-x, {\pi/2}-u)
\Pi(x, -{\pi/2}+u)\Pi(-x, -{\pi/2}+u)}\over{\Pi^{2}(x, {\pi/2})
\Pi^{2}(-x, {\pi/2})}};$$
$$\Pi(x, u) = \prod_{l=0}^{\infty}{{\Gamma(1/2 + (2l+1/2)\lambda +
{{x}/\pi} - {{\lambda u}/\pi})\Gamma(1/2 + (2l+3/2)\lambda +
{{x}/\pi})}\over{\Gamma(1/2 + (2l+3/2)\lambda +
{{x}/\pi} - {{\lambda u}/\pi})\Gamma(1/2 + (2l+1/2)\lambda +
{{x}/\pi})}} \eqno(5.23)$$
solves

$$\sigma(x,u)\sigma(x,-u) = [\cos(x + \lambda u)\cos(x -
\lambda u)]^{-1};\quad \sigma(x,{\pi/2}-u)=\sigma(x,{\pi/2}+u), \eqno(5.24)$$
and the parameters $\eta$ and $\vartheta$ are determined through the equations

$$\cos(\eta)\cosh(\vartheta) = - {1\over k}\cos\xi ; \quad \cos^2
(\eta) + \cosh^2 (\vartheta) = 1 + {1\over {k^2}}. \eqno(5.25)$$

The above boundary S-matrix has a very rich structure. The factor
$\sigma(i\vartheta,u)$ in (5.22) brings in an infinite set of singularities
at complex values of $u$; some of these complex poles probably can be
interpreted as resonance states of the boundary. The factor $\sigma(\eta,u)$
exhibits infinitely many  poles at real $u$; depending on the parameter
$\eta$, some of these poles can occur in the ``physical strip''
$0 < u < {\pi/2}$ thus giving rise to the boundary bound states.
We did not carry out a  complete analysis of the possible boundary
phenomena described by this S-matrix. In order to do this one must
find the an exact relation between the parameters $M$ and $\varphi_0$ in
the boundary action and $\xi$ and $k$ (or $\eta$ and $\vartheta$)
in the S-matrix. A brief analysis shows that in the general case this
relation is very complicated. We leave this problem for future studies.
Here we analyze only two particular cases.

a). ``Fixed'' boundary condition $\varphi(x,y)_{x=0}=\varphi_0$ can be
obtained from (5.8) in the limit $M=\infty$. In this case the
topological charge is conserved and the amplitudes $Q_{\pm}$ in (5.9)
must vanish. Hence this case corresponds to $k=0$. We have two
amplitudes, $P_{+}$ and $P_{-}$ which describe soliton scattering
processes schematically shown in Fig.25. For $k=0$, the ``resonance''
factor $\sigma(i\vartheta,u)$ disappears and the equation (5.22)
simplifies as

$$R_{1}(u) = {1\over \cos\xi} \sigma ( \xi , u), \eqno(5.26)$$
where $\xi$ is related in some way to $\varphi_0$. Obviously,
$\varphi_0 =0$ corresponds to $\xi = 0$ as at $\varphi_0 =0$ the boundary
theory respects ${\bf C}$ symmetry and we must have $P_{+} = P_{-}$. At
$\xi =0$ the function $R_{1}(u)$ does not exhibit any poles in the
physical strip (and hence there are no boundary bound states), the
nearest singularity being the double pole at $u= -{{\pi}/2\lambda}$. At
$\xi \neq 0$ (i.e. $\varphi_0 \neq 0$) this double pole splits into two
simple poles at $u = u_{\pm} =-{{\pi}/2\lambda}\pm {\xi/\lambda}$. Note
that due to the factors $\cos(\xi \pm \lambda u)$ in (5.12) the pole
$\theta=iu_{+}$ ($\theta=iu_{-}$) appears only in $P_{+}$ ($P_{-}$). At
$\xi > {\pi/2}$ the pole at $\theta=iu_{+}$ enters the physical strip.
Appearence of such a ``boundary bound state'' is well expected.
For $0 < \varphi_0 < {\pi/\beta}$ the ground state $\mid 0 \rangle_B$
of the boundary sine-Gordon theory is characterized by the asymptotic
behaviour $\langle \varphi(x,y) \rangle_B \to  0$ as $x \to -\infty$.
Classically, there is another stable state with $\langle \varphi(x,y)
\rangle \to  {2\pi/\beta}$ as $x \to -\infty$. If $\varphi_0$ is small
(compared to the quantum parameter $\beta$), quantum fluctuations can
destroy its stability.  However, if $\varphi_0$ is not too small, this
state is expected to be stable in the quantum theory, too. At $\varphi_0 =
{\pi/ \beta}$  this state has the same energy as the ground state which
hence becomes degenerate. In this case the amplitude $P_{+}(\theta)$
has to have a pole at $\theta = i{\pi/2}$ corresponding to the ``emission''
of a zero-momentum soliton by the associated boundary state, as is
explained in Sect.3. We can conclude that $\varphi_0={\pi/ \beta}$
corresponds to the case $u_{+} = {\pi/2}$. Assuming a linear relation
between $\xi$ and $\varphi_0$, we find

$$\xi = {{4\pi}\over \beta}\varphi_0. \eqno(5.27)$$
Let us stress that the assumed linear relation (and hence (5.27))
is no more than a conjecture. Also, the relation (5.27) is conjectured
only for the ``fixed'' case $M=\infty$; it is not expected to hold in
the general case of finite $M$.

b). ``Free'' boundary condition corresponds to $M=0$ in (5.8). In this
case the full action enjoys the symmetry $\varphi \to -\varphi$ ($\bf C$
symmetry) and hence we must have $P_{+}=P_{-}=P_{free}$ and
$Q_{+}=Q_{-}=Q_{free}$, i.e.
we have to choose $\xi = 0$. Note that under this choice the factors
$\cos(\lambda \theta)$ in (5.12) cancel the poles at $\theta =
iu_{2n+1}$ of the factor (5.21). This agrees with the expectation that
in the $\bf C$ -symmetric case the boundary state can not emit the $\bf C$
-odd particles $B_{2n+1}$\footnote{$^3$}{In fact, it is easy to argue
that in general the choice $\xi =0$ corresponds to the $\bf C$ -symmetric
boundary action (5.8) with $\varphi_0 =0$. Even in this case the
relation between $M$ and $k$ seems to be rather complicated.}. In
addition, in the ``free'' case one expects all the amplitudes (5.12) to
have a pole at $\theta = {i\pi/2}$, for exactly the same reason which we
discussed in the previous Section. To ensure this property one has to
choose $\eta = {\pi\over 2}(\lambda + 1)$ (and $\vartheta = 0$), i.e. $k =
[{\sin({\pi\lambda}/2)}]^{-1}$. So in the ``free'' case the amplitudes
(5.12) can be written as

$$P_{free}(\theta) = \sin({\pi\lambda}/2)R_{free}(u); \qquad
Q_{free}(\theta) = \sin(\lambda u)R_{free}(u) \eqno(5.28)$$
where

$$R_{free}(u)={{\cos(\lambda u)}\over {\sin({\lambda\pi}/2)}}
R_{0}(u)\sigma({{4\pi^2}/{\beta^2}},u)\sigma(0,u).\eqno(5.29)$$

\vskip 0.4in

6.DISCUSSION

Obviously, there is much room for further development.

More detailed analysis of ``integrable boundary conditions'' is needed.
In Section 2 we assumed that the boundary action $b$ in (2.8) depends
only on boundary values $\varphi_B (y), {d/dy}\varphi_B (y)$ of the
``bulk'' field $\varphi(x,y)$. In general, the boundary action can contain
also specific boundary degrees of freedom. Clearly, this provides more
possibilities for the integrable boundary field theories.

Further study of possible solutions to Boundary Yang-Baxter equations
((3.25) is of much interest. A new class of solutions is found in a recent
paper [24].

Many integrable ``bulk'' theories exhibit nonlocal integrals of motion
of fractional spin [25-27], along with the local ones. Analysis of these
nonlocal integrals provides valuable information about the S-matrix. We
expect that the nonlocal integrals of motion exist in some integrable
boundary field theories. A natural place to look for the nonlocal IM is
boundary sine-Gordon model considered in Section 5. As in the case of
bulk sine-Gordon theory, where nonlocal IM generate the so-called ``affine
quantum group symmetry''[27], appropriate modifications of these IM could
help to find an exact relation between the parameters of the action ($M$
and $\varphi_0$ in (5.8)) and of the S-matrix ($\xi, k$ in (5.12))(which
we found very difficult to guess).

The most interesting problem is how to extract any off-shell data from
the S-matrix. In the bulk theory, two approaches have proven to be the most
successfull. One is the ``formfactor bootstrap'' proposed by Smirnov[27]. It
would be interesting to generalize the formfactor bootstrap equations to the
case of integrable boundary field theory, both for bulk operators in
presence of boundary and for boundary operators. Another approach is
known as ``thermodinamic Bethe anzats''(TBA)[29]. At the present stage of
development, it is capable of providing the ground state energy of a
finite-size system (with spatial coordinate compactified on a circle),
once the S-matrix is known. This approach can be generalized in different
ways to incorporate the boundary effects. In particular, the expression
(3.38) for the boundary state allows one to find the analog of TBA equations
for spatially finite systems with integrable boundaries. We intend to
study these equations elsewhere.

It was understood recently that many interesting ``massless flows'' in
the bulk theory can be described in terms of ``massless factorizable
S-matrices'', through TBA equations[30]. This approach can be modified to
incorporate ``massless boundary S-matrices''\break [31], thus providing the
possibility
of a similar description of the ``boundary flows''[32].

After this work was finished we received a recent paper [31] where some
results of Section 3 (notably, the ``boundary bootstrap
equation''(3.40)) are obtained. However, the ``boundary cross-unitarity
equation'' (3.35) (which we think to be the most significant of our
results) is not considered there; that is why we decided that our paper
is still worth publishing.

\vskip 0.4in

APPENDIX

Here we carry out a preliminary study of integrable boundary conditions in
the sine-Gordon model (5.1). We restrict attention to the classical case $\beta
\to 0$. In this limit it is convenient to work with the field $\phi(x,y)
= \beta\varphi(x,y)$ so that the ``bulk'' equation of motion takes the
form

$$\partial_{z}\partial_{\bar z}\phi = \sin\phi, \eqno(A.1)$$
where we set $m^2 = 4$. As is known [22], there are infinitely many
conserved currents $(T_{s+1},\Theta_{s-1})$ and $(\bar T_{s+1},
\bar\Theta_{s-1})$ with $s = 1, 3, 5, ...,$ which satisfy (2.5) in this
``bulk'' theory; the first two have the form

$$T_{2} = (\partial_{z}\phi)^2; \qquad \Theta_0 = -2\cos\phi, \eqno(A.2)$$
$$T_{4} = (\partial_{z}^{2}\phi)^2 - {1\over 4}(\partial_{z}\phi)^{4};
\qquad \Theta_{2} = (\partial_{z}\phi)^{2}\cos\phi \eqno(A.3)$$
($\bar T$ and $\bar\Theta$ are obtained from (A.2) by substitution $z
\to \bar z$). The current (A.2) is just the energy-momentum tensor, and
(A.3) leads to the first nontrivial IM.

As is explained in Section 2, the current (A.3) would lead to a
nontrivial IM in presence of the boundary, too, provided the boundary
condition at $x=0$ is chosen in such a way that

$$[T_4 + \bar \Theta_2 - \bar T_4 - \Theta_2]_{x=0} = {d\over
{dy}}\theta_3, \eqno(A.4)$$
where $\theta_3$ is some local boundary field. Let us consider the
simplest case of the boundary action $b$ independant of the derivative
${d\over dy}\phi$, i.e. the total action of the form

$${\cal A}={1\over {\beta^2}}\lbrace\int_{-\infty}^{\infty}dy
\int_{-\infty}^{0}dx [{1\over 2}(\partial_{\mu}\phi(x,y))^2 -
4\cos\phi(x,y)] + \int_{-\infty}^{\infty}dy V(\phi(x=0,y))\rbrace.
\eqno(A.5)$$
The boundary condition which follows from (A.5) is

$$[\partial_{x}\phi + V'(\phi)]_{x=0} = 0. \eqno(A.6)$$
With this, the l.h.s of (A.4) can be written (up to an overall numerical
factor) as

$$V'(\phi)(\partial_{y}\phi)^3 +
8V''(\phi)(\partial_{yy}^2 \phi)(\partial_{y}\phi) - [(V'(\phi))^3 +
4V''(\phi)\sin\phi + 2V'(\phi)\cos\phi](\partial_{y}\phi). \eqno(A.7)$$
This reduces to a total $d/dy$ derivative if the function $V(\phi)$
satisfies

$$4V''(\phi) + V(\phi) = 0, \eqno(A.8)$$
that is

$$V(\phi) = -\Lambda \cos({{\phi - \phi_0}\over 2}), \eqno(A.9)$$
where $\Lambda$ and $\phi_0$ are constants. Returning to the original
normalization, one gets (5.8).

\vfil\break

REFERENCES.

1.K.Binder, in ``Phase Transitions and Critical Phenomena'', v.8, C.Domb

and J.Lebowitz ed., Academic Press, London, 1983.

2.C.G.Callan, L.Thorlacius. Nucl.Phys.B329(1990),117.

3.E.Witten. Phys.Rev.46D(1992)p.5467., K.Li, E.Witten. Preprint

IASSNS-HEP-93/7, 1993.

4.A.B.Zamolodchikov, Al.B.Zamolodchikov. Ann.Phys.120 (1979), 253.

5.A.B.Zamolodchikov. Advanced Studies in Pure Mathematics 19 (1989), 641.

6.J.Cardy. ``Conformal Invariance and Surface Critical Behavior'', in

``Conformal Invariance and Applications to Statistical Mechanics'',

E.Itzykson, H.Saleur, J.B.Zuber eds., World Scientific, 1988.

7.J.Cardy. Nucl.Phys.B324(1989)p. 581.

8.J.Cardy, D.Lewellen. Phys.Lett. 259B (1991) p.274.

9.E.Ogievetsky, N.Reshetikhin, P.Wiegmann. Nucl.Phys B280 [FS 18]
(1987), p.45.

10.V.Fateev, A.Zamolodchikov. ``Conformal Field Theory and Purely

Elastic S-Matrices'', in ``Physics and Mathematics of Strings'',

memorial volume for Vadim Knizhnik, L.Brink, D.Friedan, A.Polyakov

eds., World Scientific, 1989.

11.J.Cardy, G.Mussardo. Phys. Lett. B225 (1989) 243.

12.P.Freund, T.Classen, E.Melzer. Phys.Lett. 229B (1989) p.243;

G.Sotkov, C.Zhu. Phys.Lett. 229B (1989) p.391; P.Christe, G.Mussardo.

Nucl.Phys. B330 (1990) p.465.

13.I.Cherednik. Theor.Math.Phys., 61, 35 (1984) p.977.

14.L.Takhtadjian, L.Faddeev. Usp.Mat.Nauk. 34 (1979) p.13.

15.E.Sklyanin. J.Phys. A:Math.Gen. 21 (1988) p.2375

16.L.Mezincescu, R.Nepomechie. Int.J.Mod.Phys. A6, 5231 (1991).

17.P.Kulish, E.Sklyanin. ``Algebraic Structures related to Reflection

Equation''. Preprint YITP/K-980, 1992.

18.J.Cardy. Phys.Rev.Lett. 54 (1985) p.1354.

19.M.Sato, T.Miwa, M.Jimbo. Proc. Japan Acad., 53A (1977), 6.

20.I.Affleck, A.Ludwig. ``Universal Non-integer Groundstate Degeneracy

in Critical Quantum Systems''. Preprint UBCTP-91-007, 1991.

21.B.McCoy, T.T.Wu. ``The two-dimensional Ising Model'', Harward

University Press. 1973.

22.L.Takhtadjian, L.Faddeev. Theor.Math.Phys. 21, (1974) p.160.,

V.Korepin, L.Faddeev. Theor.Math.Phys. 25 (1975) p.147.

23.S.Coleman. Phys.Rev. D11, (1975) p.2088.

24.H.J.DeVega, A.Gonzalez Ruiz. Preprint LPTHE 92-45.

25.M.Luscher. Nucl.Phys. B135 (1978) p.1.

26.A.B.Zamolodchikov.''Fractional-Spin Integrals of Motion in Perturbed

Conformal Field Theory'', in ``Fields, Strings and Quantum Gravity'',

H.Guo, Z.Qiu, H.Tye eds., Gordon and Breach, 1989.

27.D.Bernard, A.LeClaire. Commun.Math.Phys., 142 (1991) p.99.

28.F.Smirnov. ``Formfactors in Completely Integrable Models of QFT'',

Adv. Series in Math.Phys.14, World Scientific, 1992.

29.Al.Zamolodchikov. Nucl.Phys. B342 (1990) p.695.

30.Al.Zamolodchikov. Nucl.Phys. B385 (1991) p.619., A.Zamolodchikov,

Al.Zamolodchikov. Nucl.Phys. B379 (1992) p.602.

31.P.Fendley. ``Kinks in the Kondo Problem''. Preprint BUHEP-93-10.

32.A.Fring, R.Koberle. ``Factorized Scattering in the Presence of

Reflecting Boundaries''. Preprint USP-IFQSC/TH/93-06, 1993.

\vfil\break
\centerline{\hbox{\psfig{figure=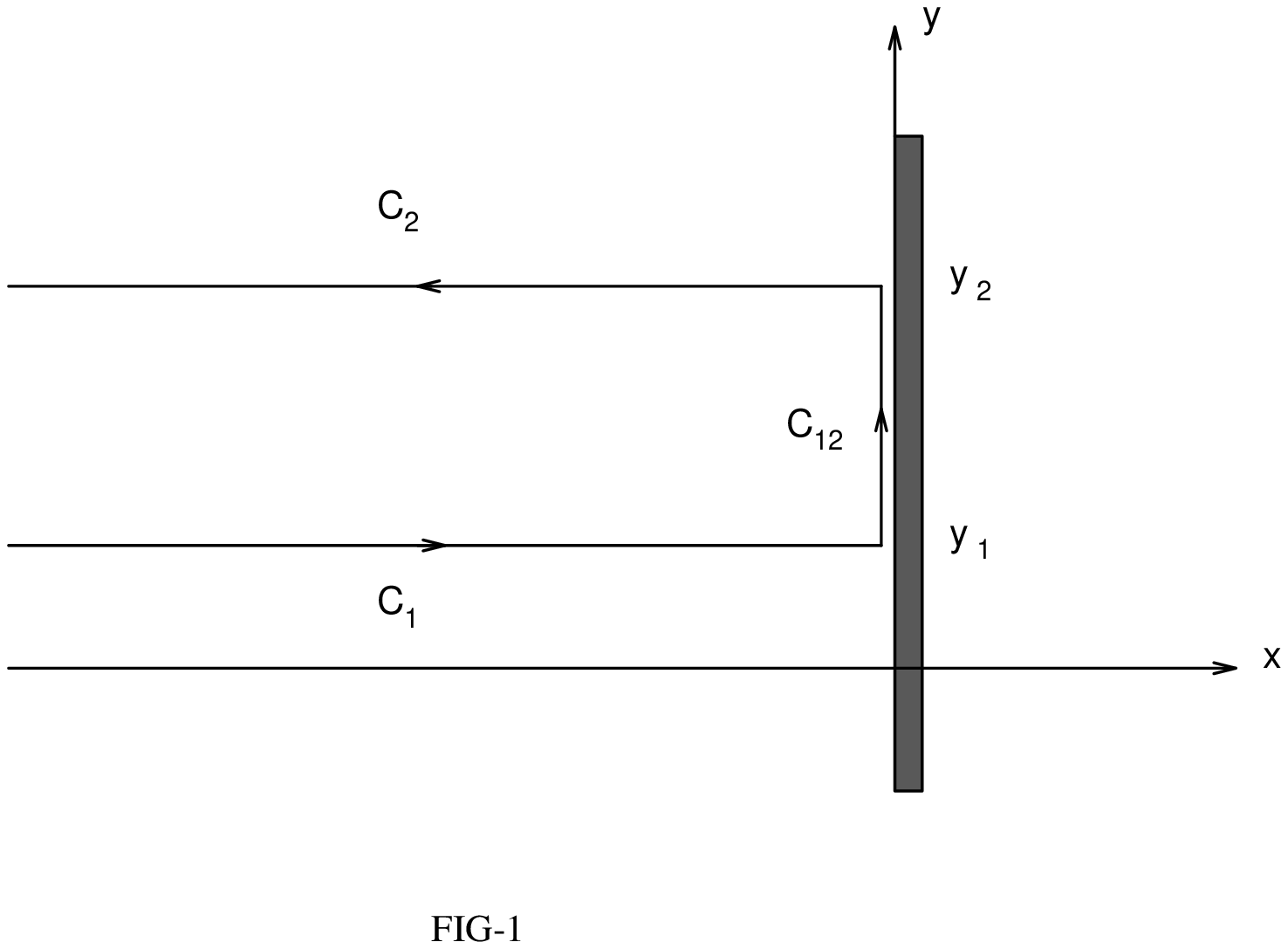,width=6in}}}
\vfil\nobreak
\centerline{\hbox{\psfig{figure=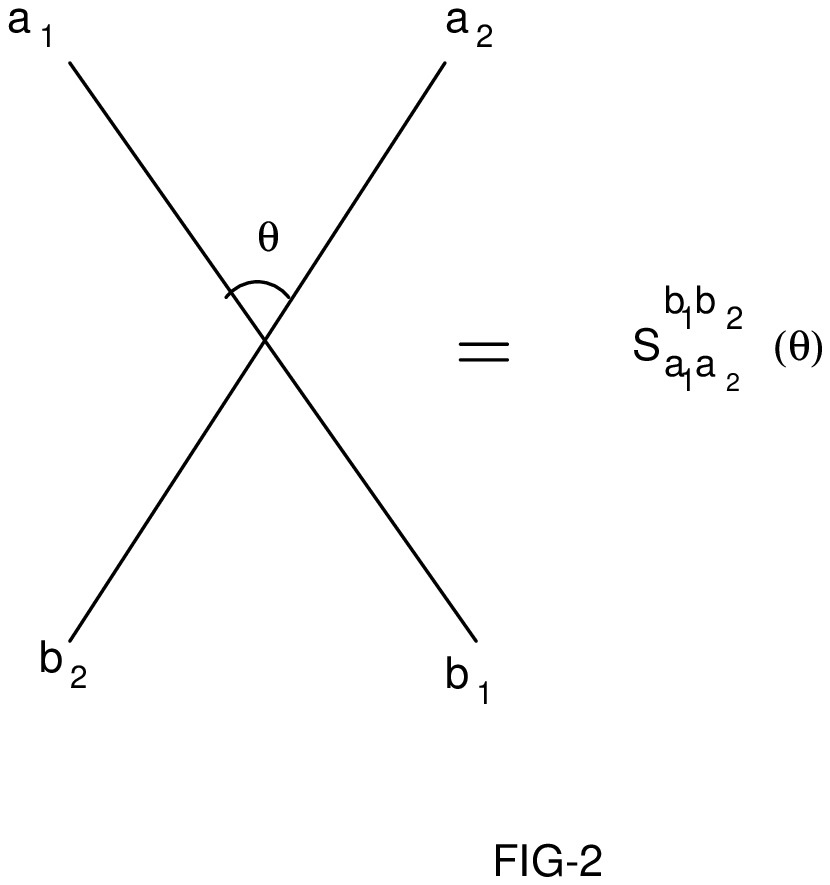,width=3.0in}\hskip0.5in
		  \psfig{figure=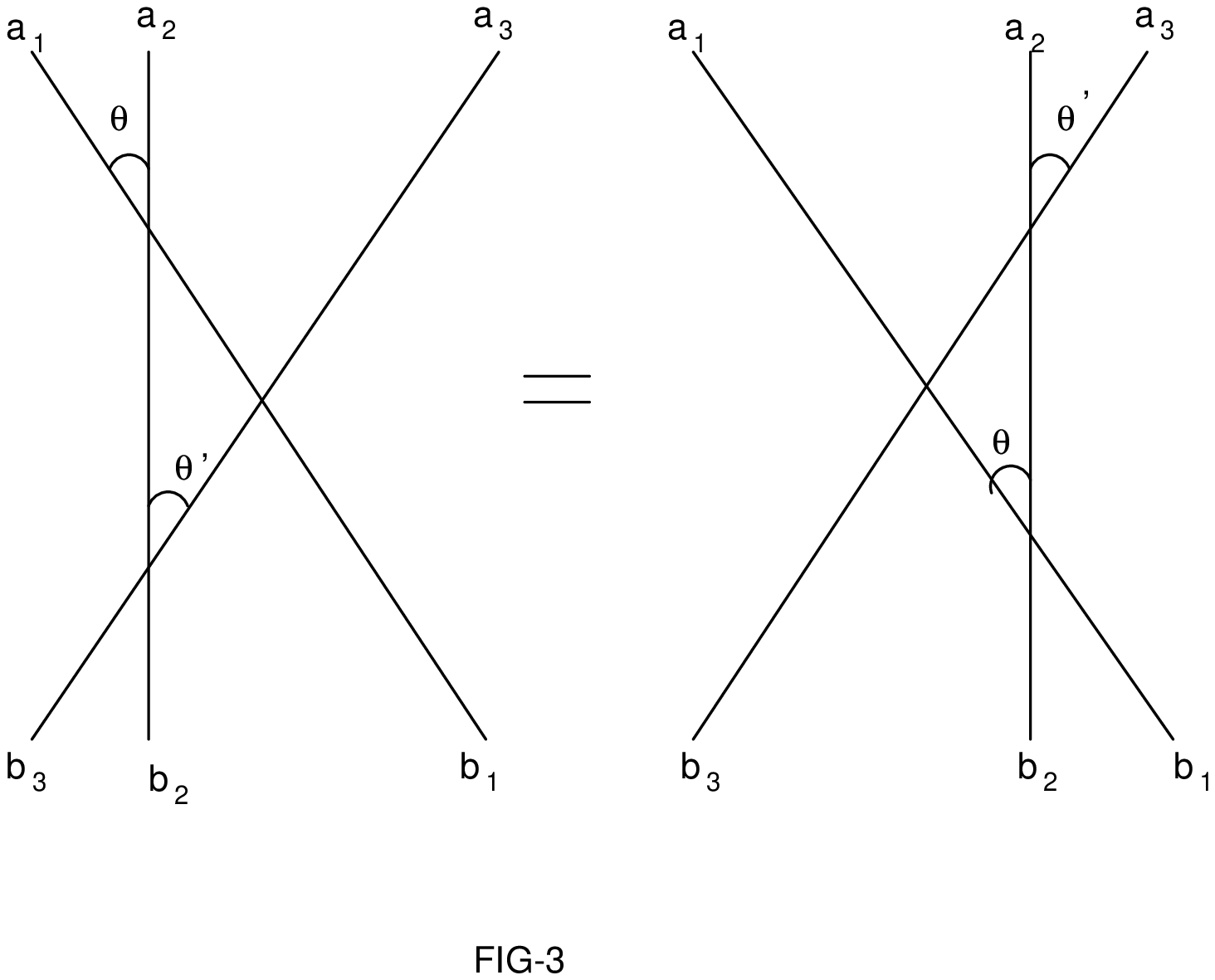,width=4.0in}}}
\break
\centerline{\hbox{\psfig{figure=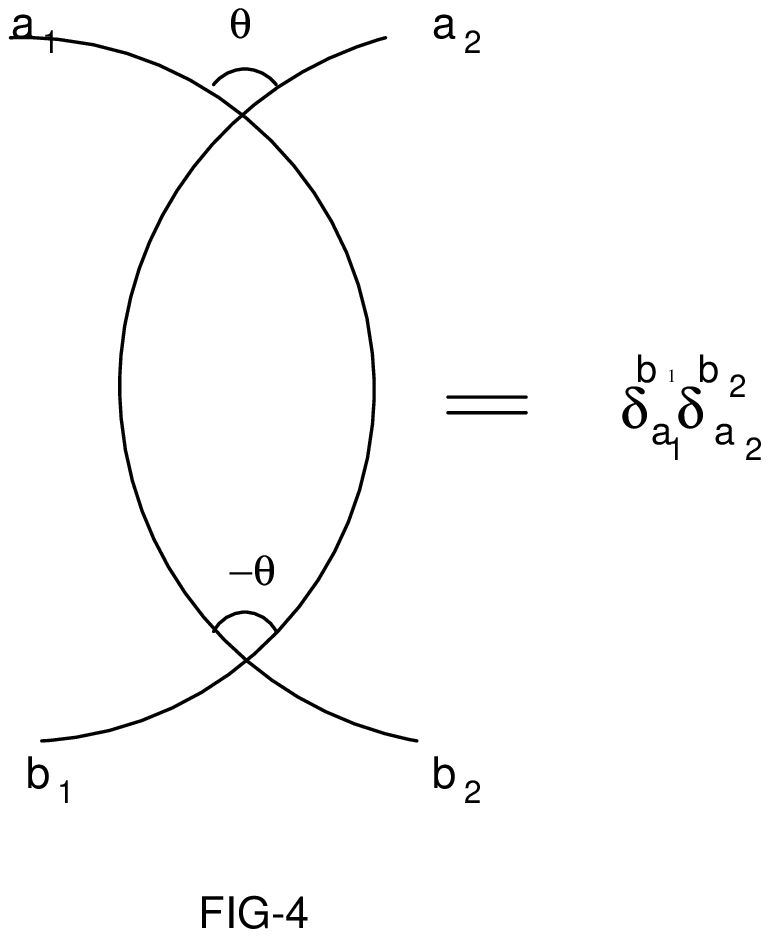,width=3.1in}\hskip1.5in
		  \psfig{figure=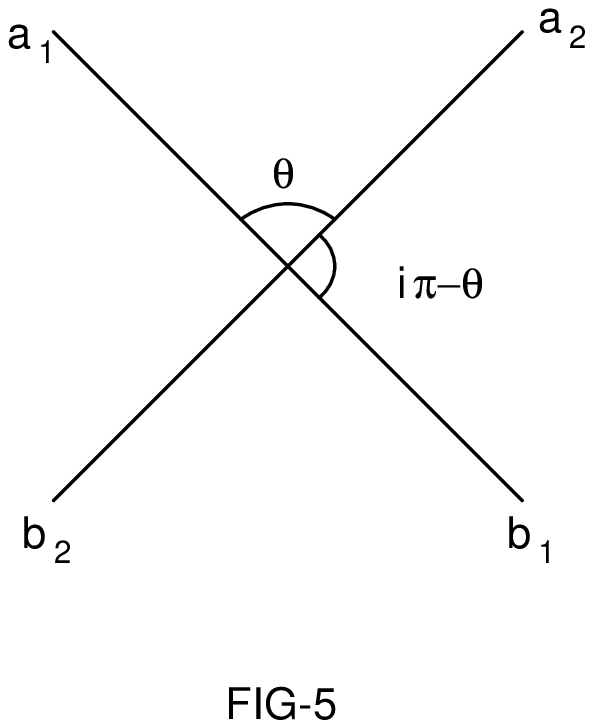,width=2.33in}}}
\vfil\nobreak
\centerline{\hbox{\psfig{figure=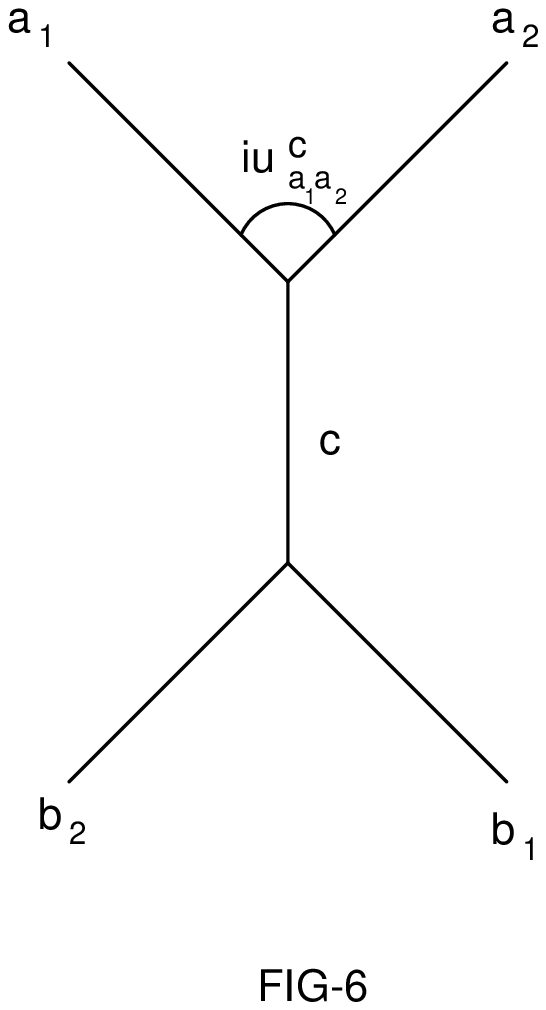,width=2.2in}\hskip1.5in
		  \psfig{figure=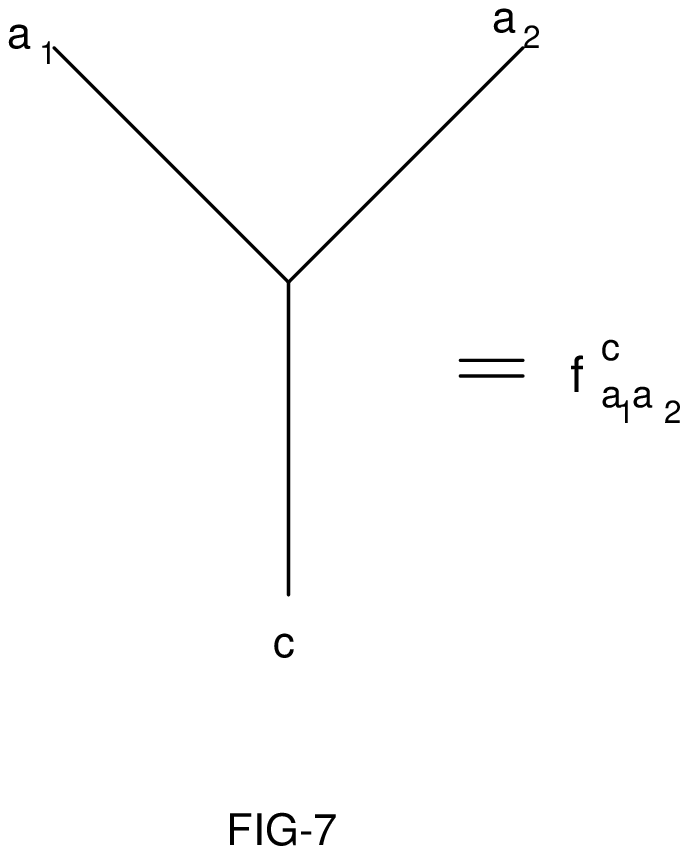,width=2.71in}}}
\break
\centerline{\hbox{\psfig{figure=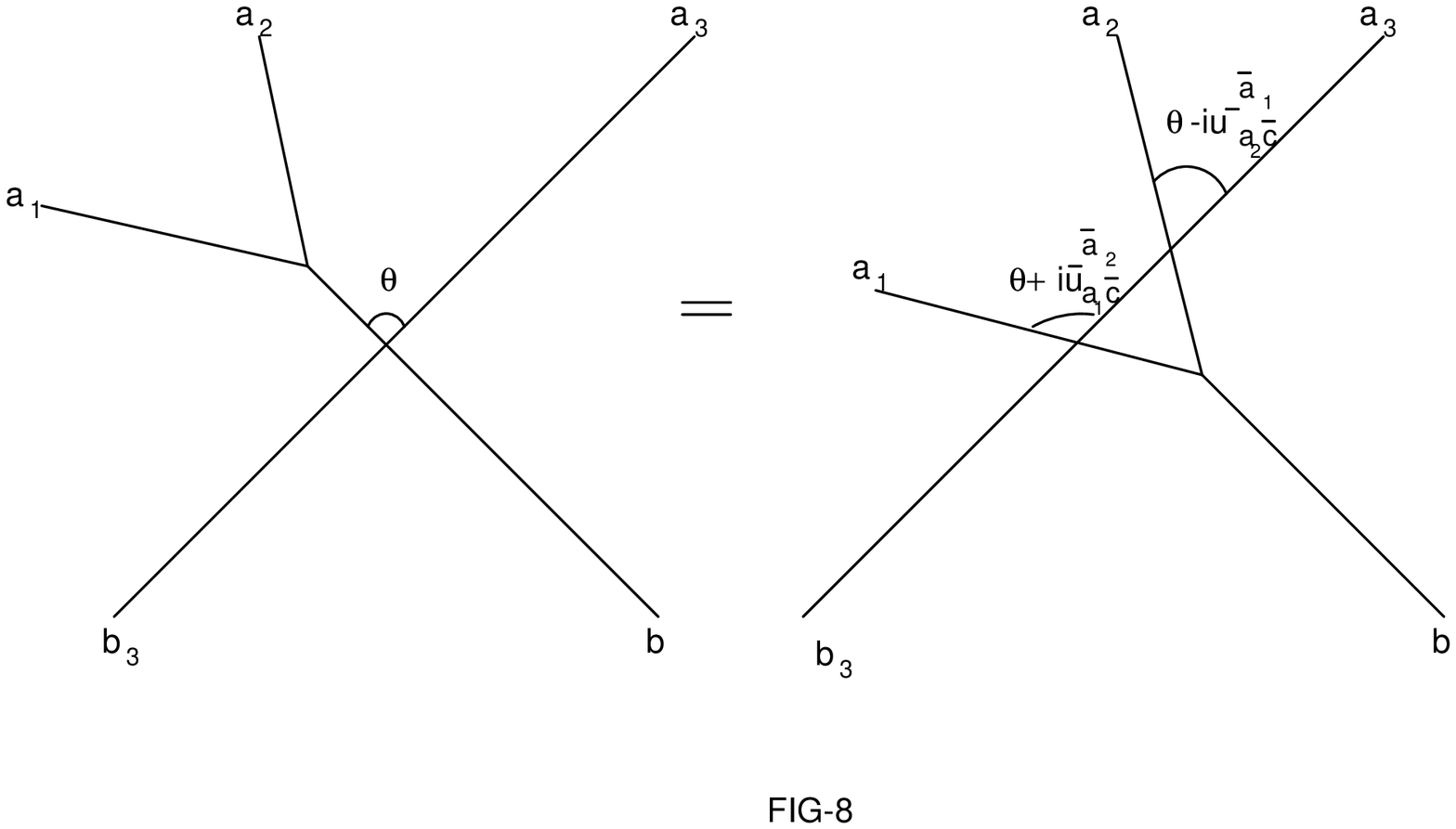,width=7.5in}}}
\vfil\nobreak
\centerline{\hbox{\psfig{figure=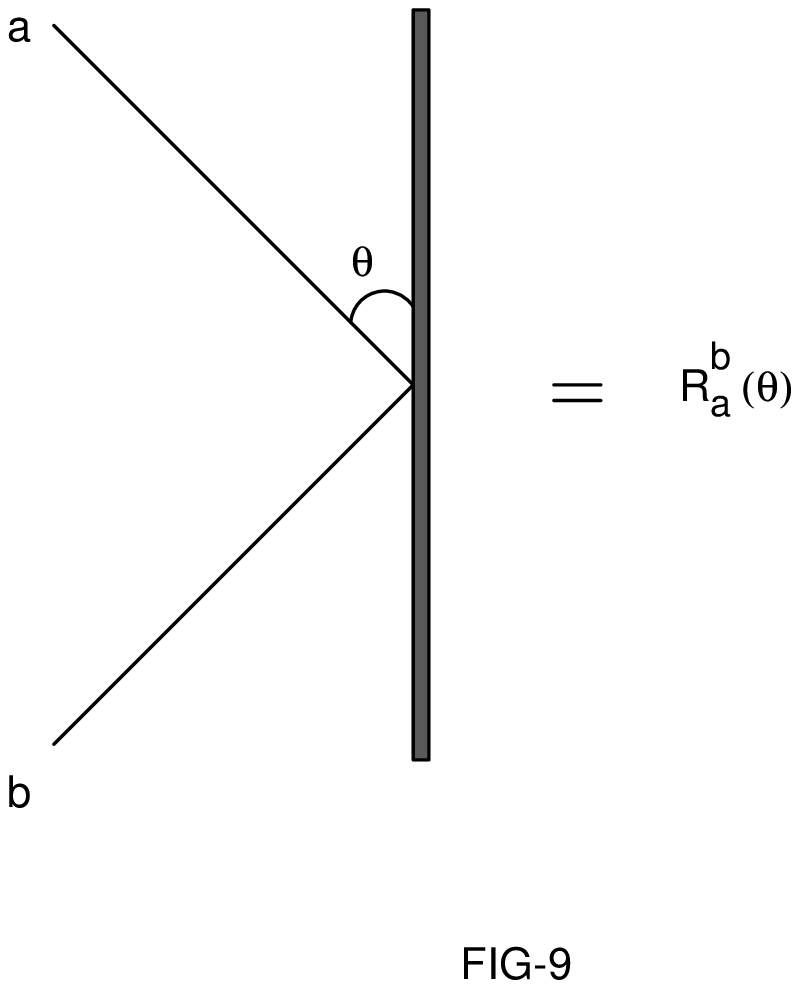,width=3.2in}\hskip1.5in
		  \psfig{figure=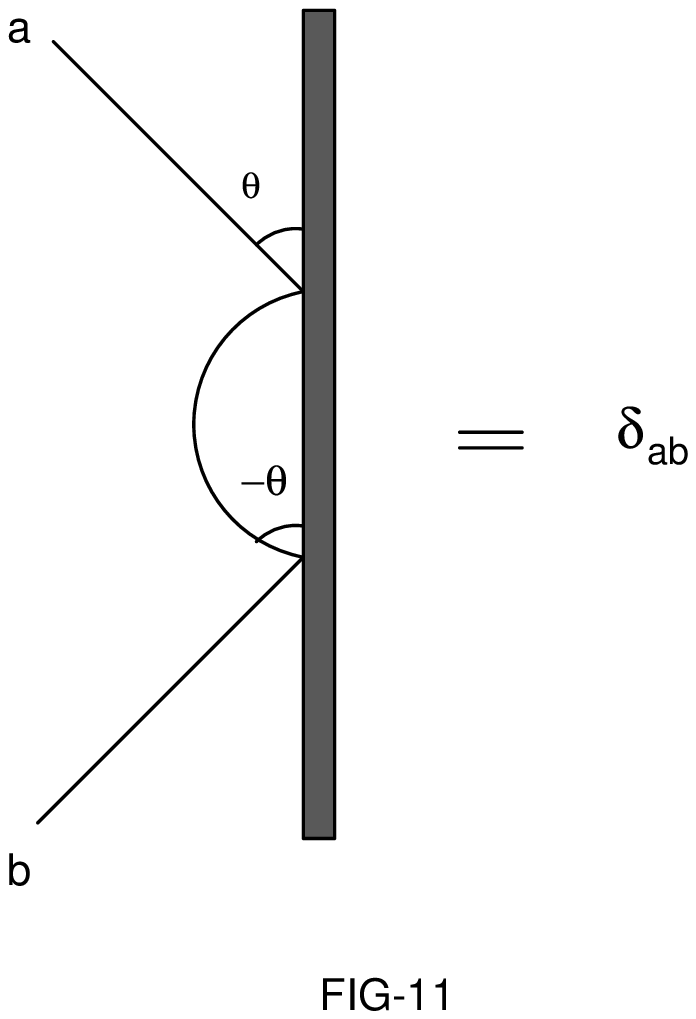,width=2.75in}}}
\break
\centerline{\hbox{\psfig{figure=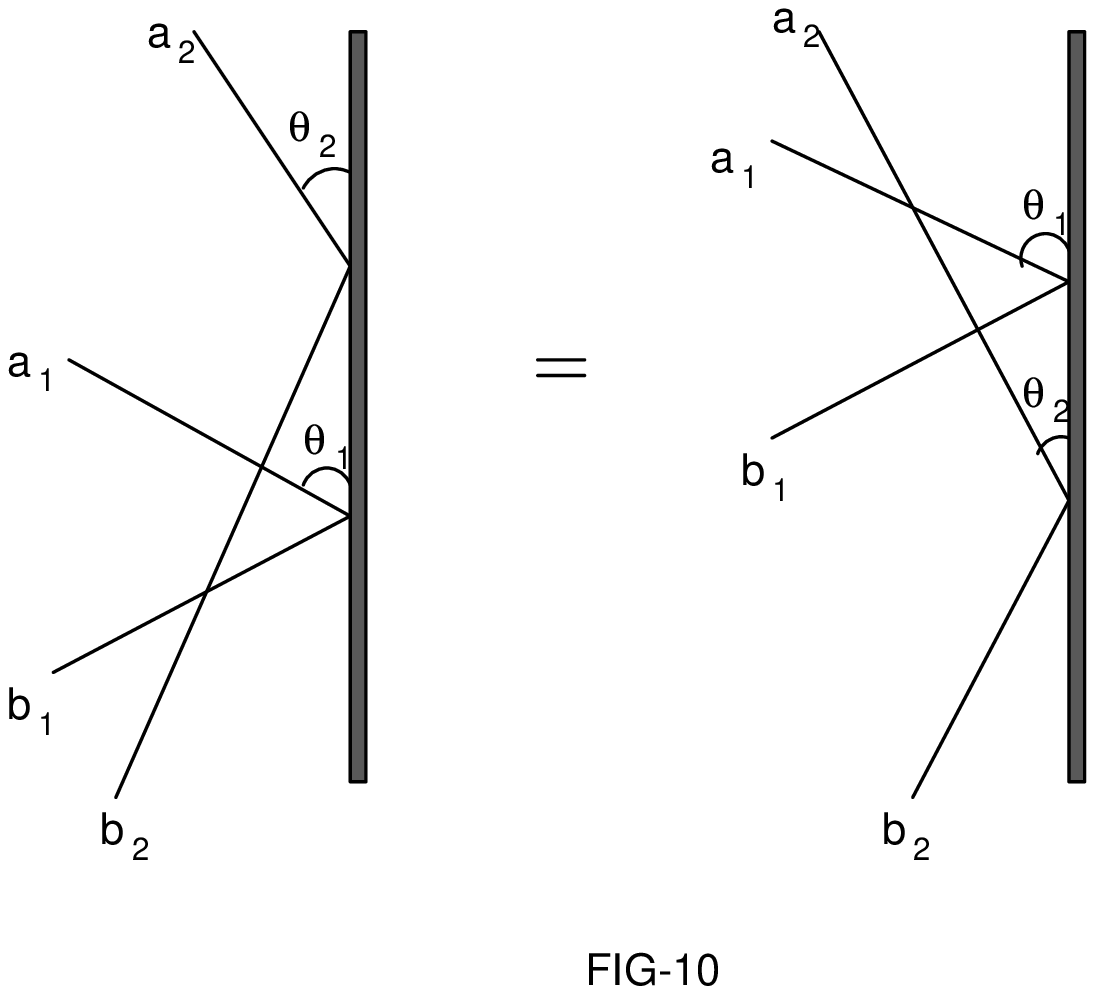,width=4.33in}}}
\vfil\nobreak
\centerline{\hbox{\psfig{figure=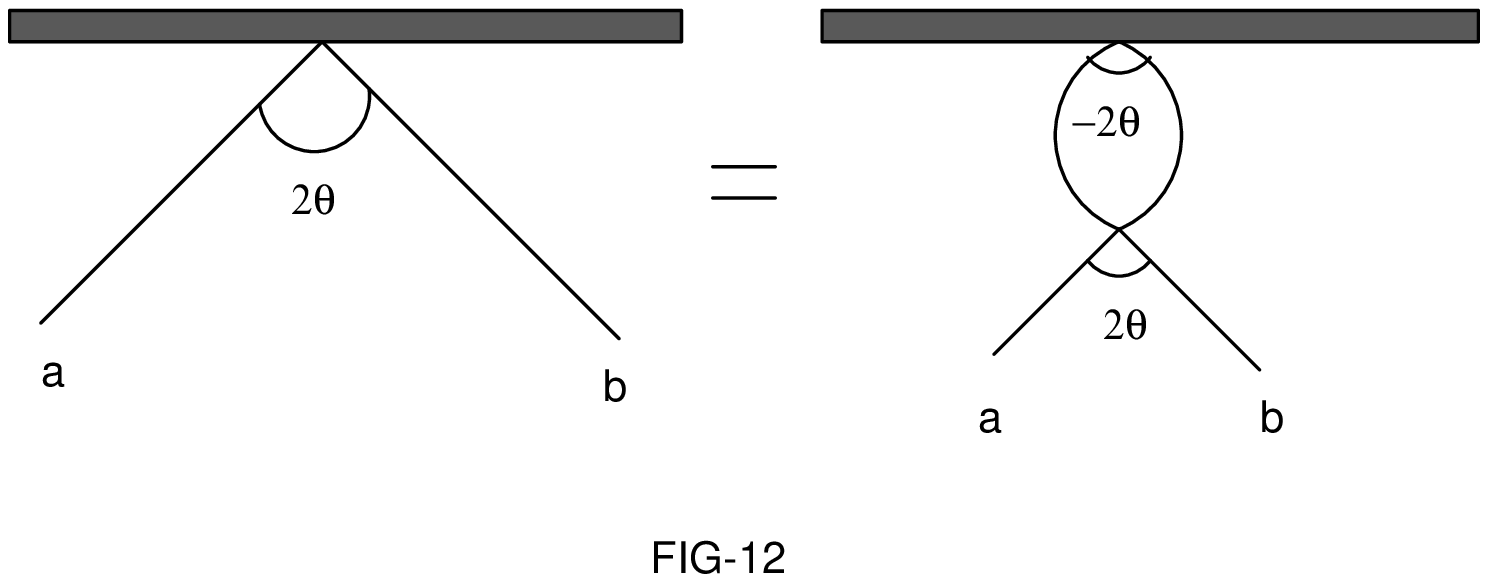,width=6.0in}}}
\break
\centerline{\hbox{\psfig{figure=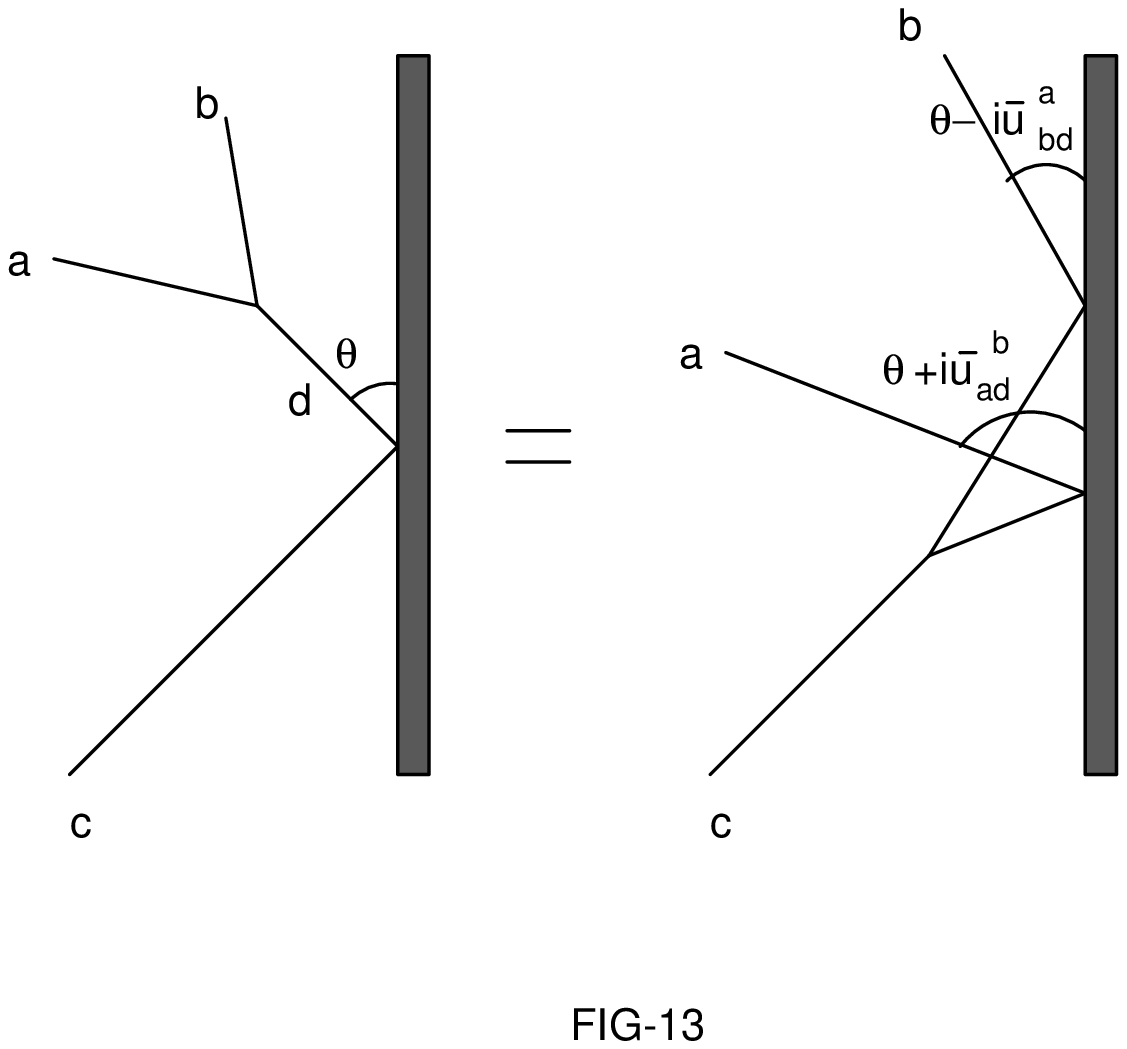,width=4.0in}\hskip0.5in
		  \psfig{figure=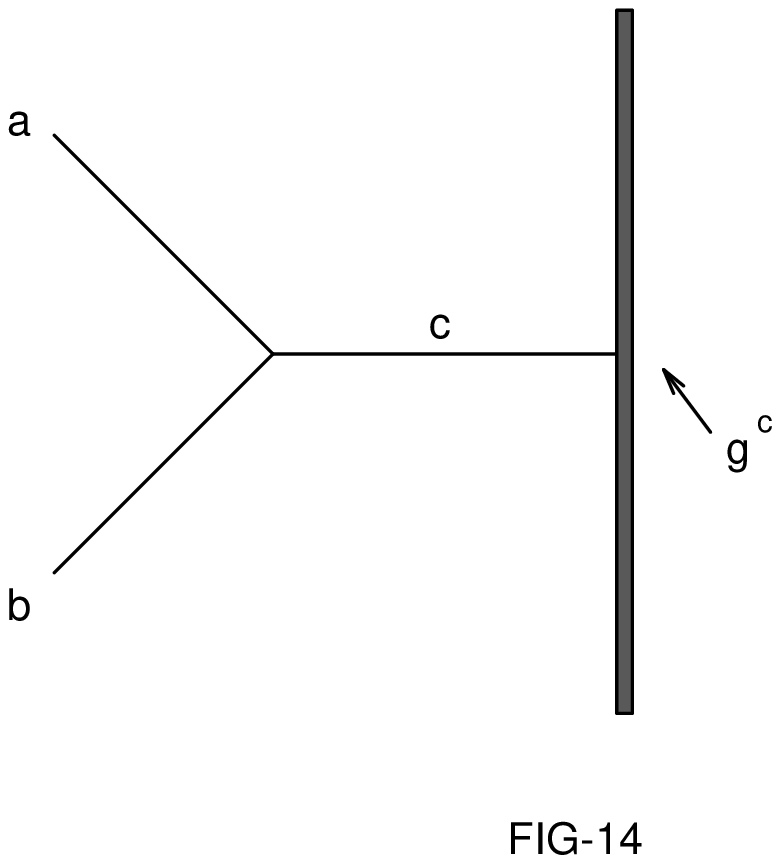,width=3.1in}}}
\vfil\nobreak
\centerline{\hbox{\psfig{figure=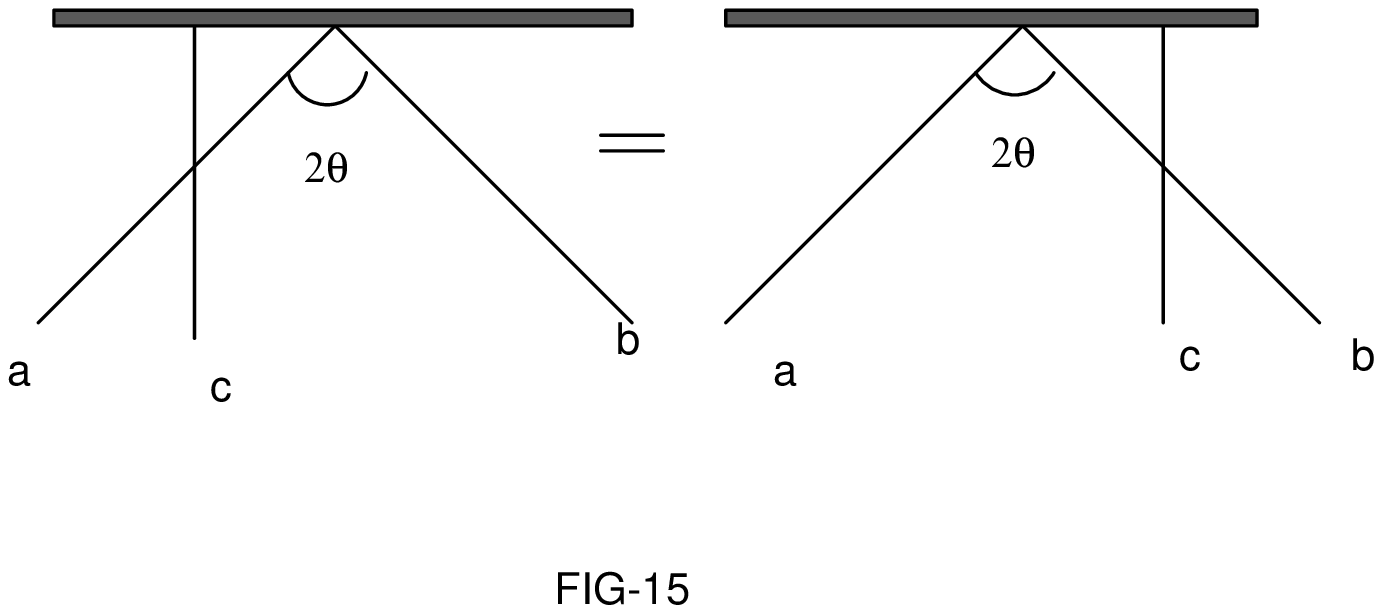,width=5.5in}}}
\break
\centerline{\hbox{\psfig{figure=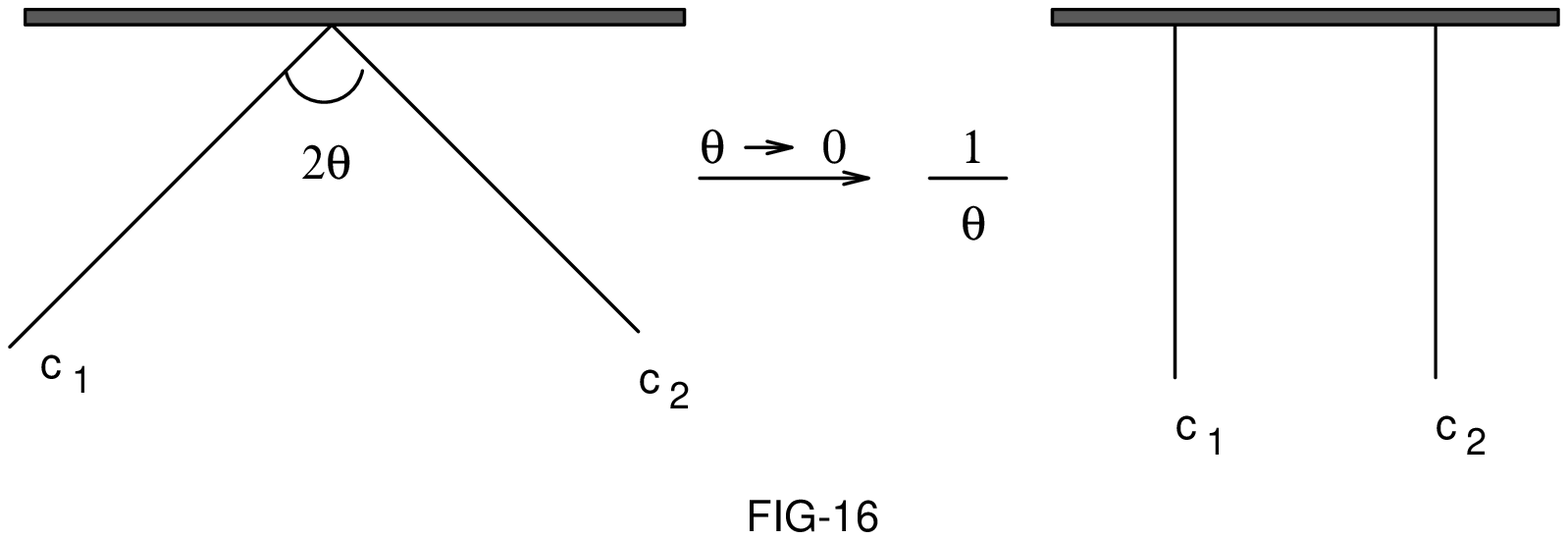,width=6.32in}}}
\vfil\nobreak
\centerline{\hbox{\psfig{figure=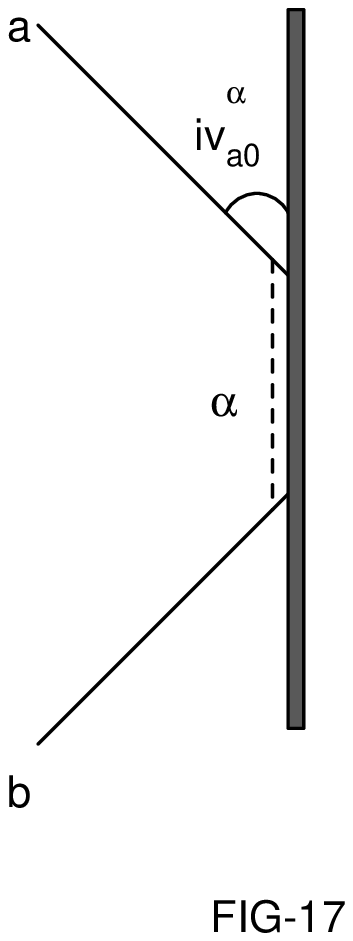,width=1.36in}\hskip1.5in
		  \psfig{figure=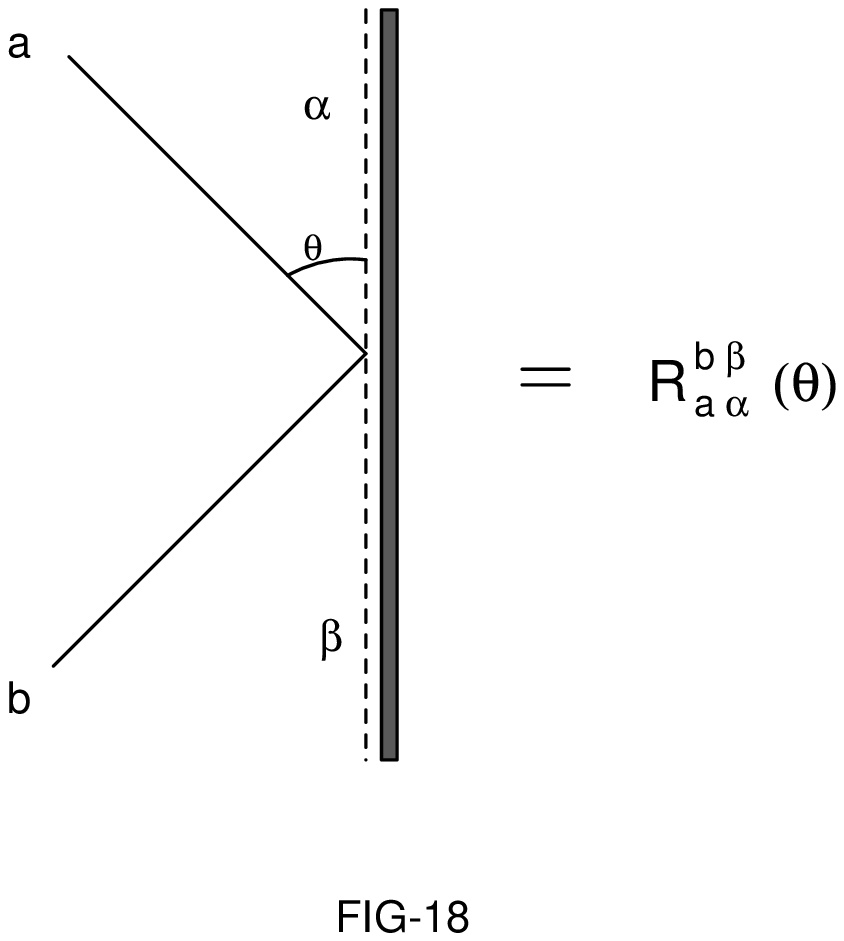,width=3.34in}}}
\break
\centerline{\hbox{\psfig{figure=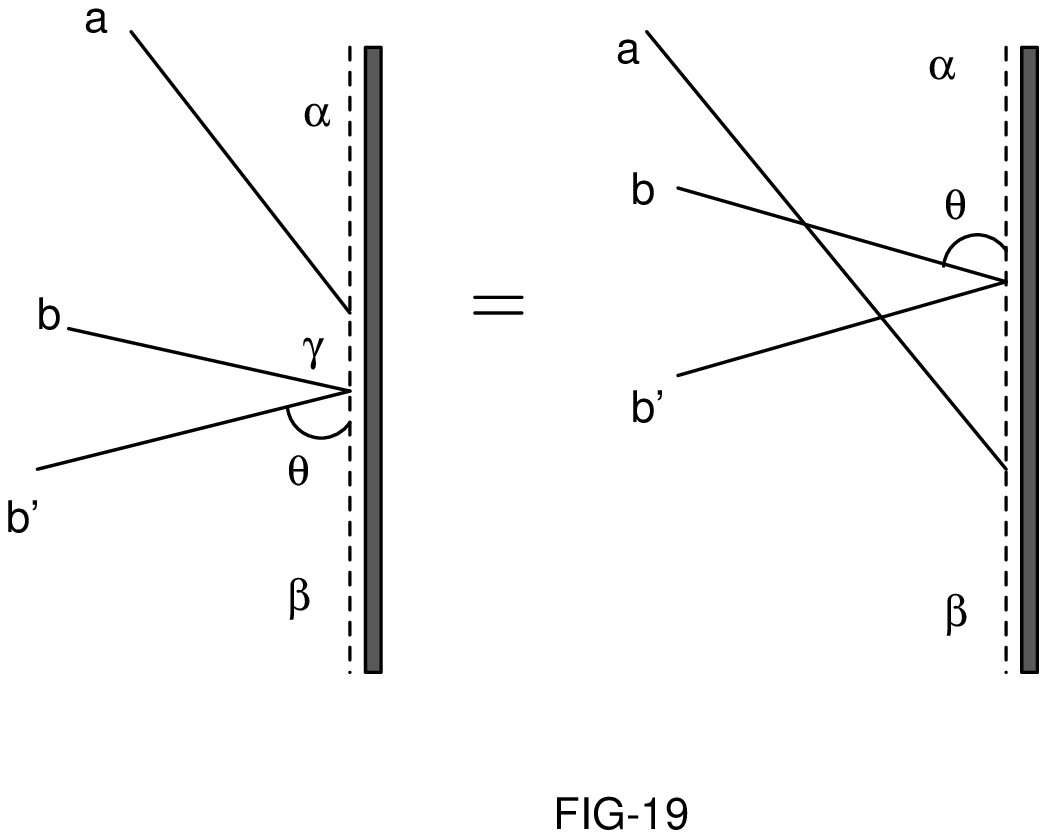,width=4.15in}}}
\vfil\nobreak
\centerline{\hbox{\psfig{figure=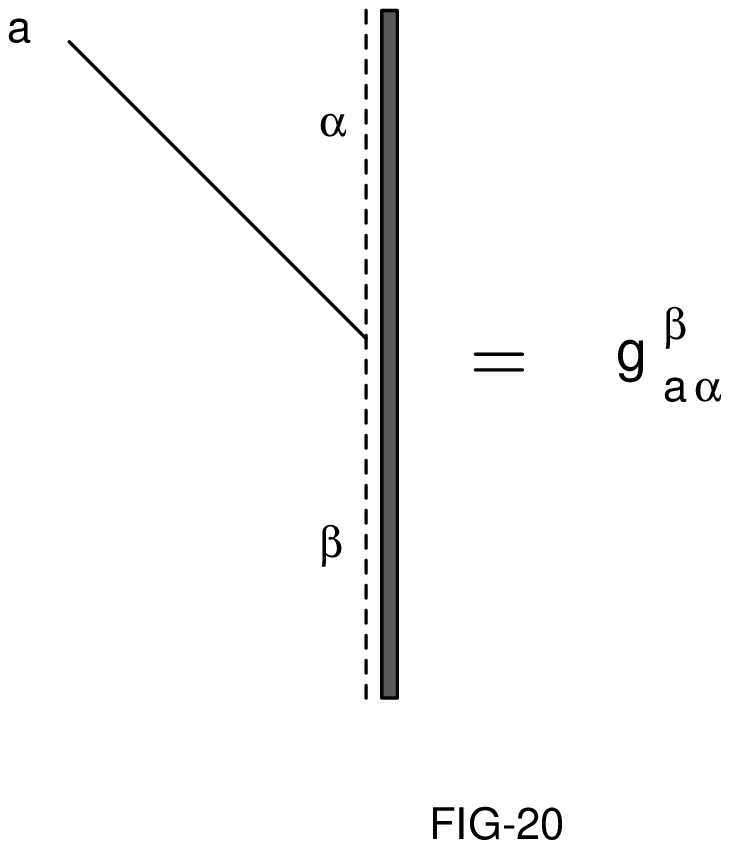,width=2.88in}\hskip1.5in
		  \psfig{figure=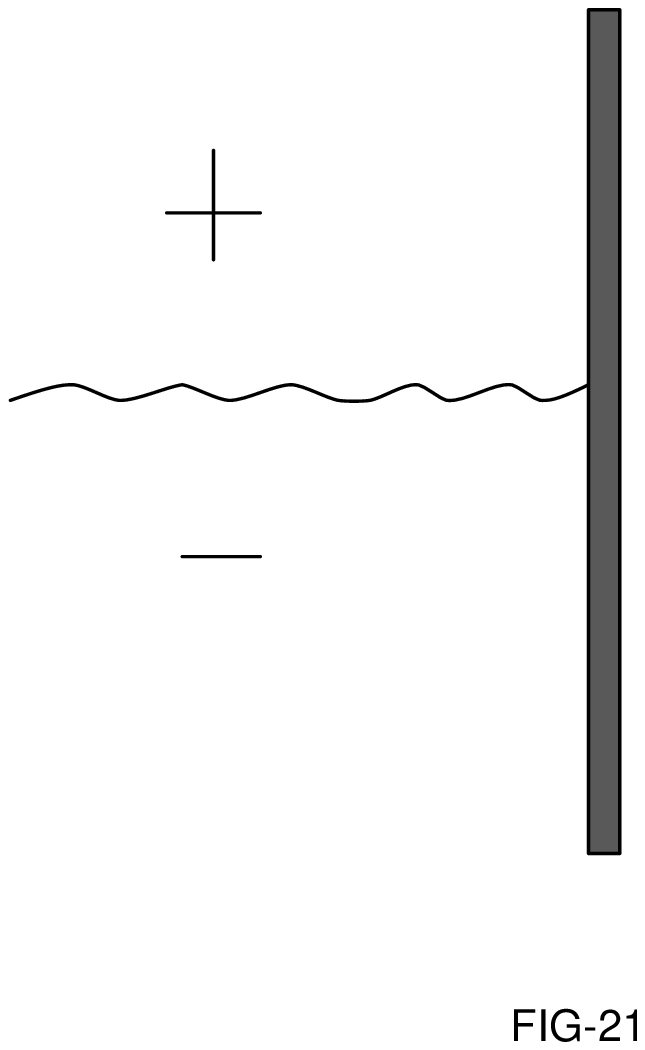,width=2.6in}}}
\break
\centerline{\hbox{\psfig{figure=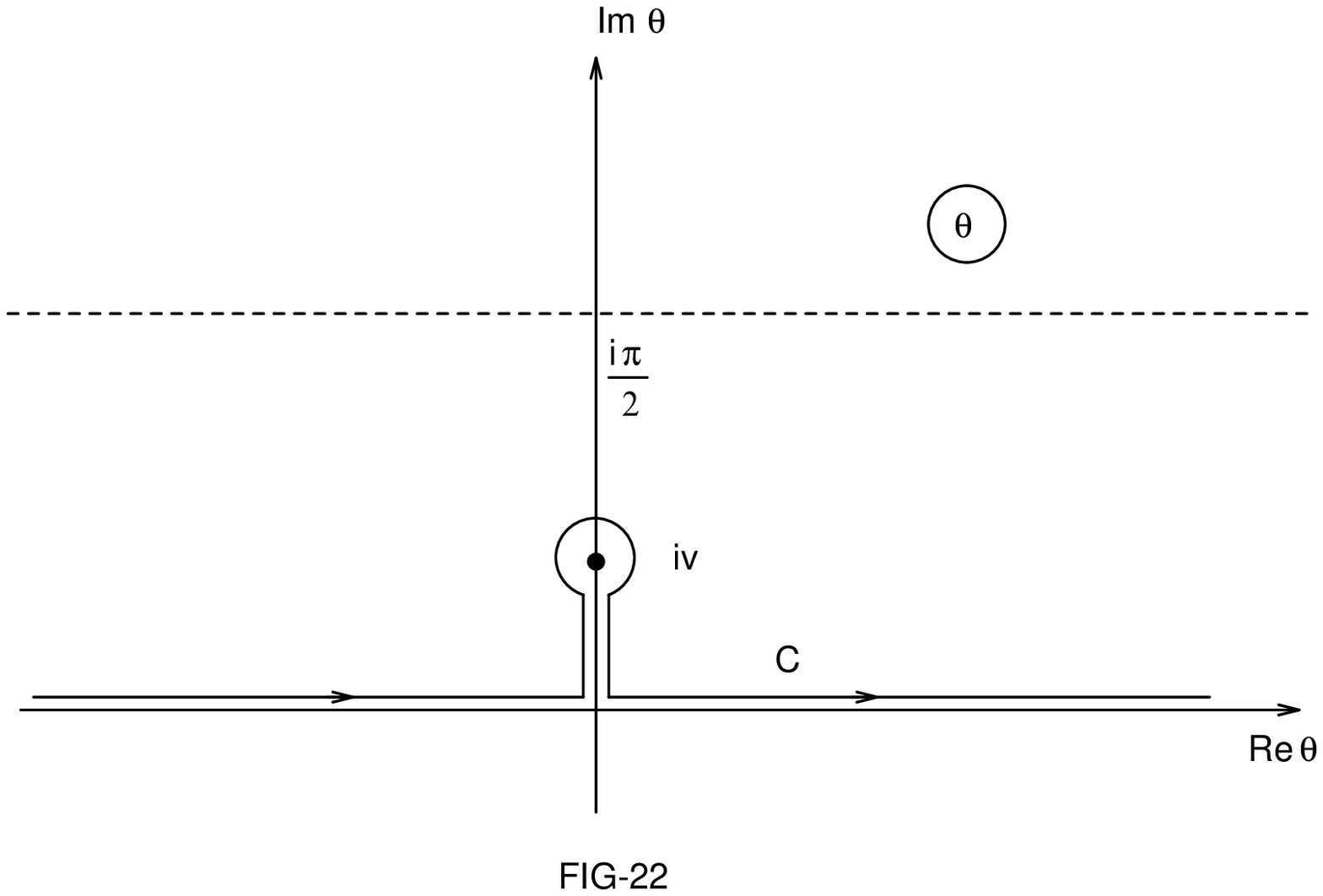,width=7.42in}}}
\vfil\nobreak
\centerline{\hbox{\psfig{figure=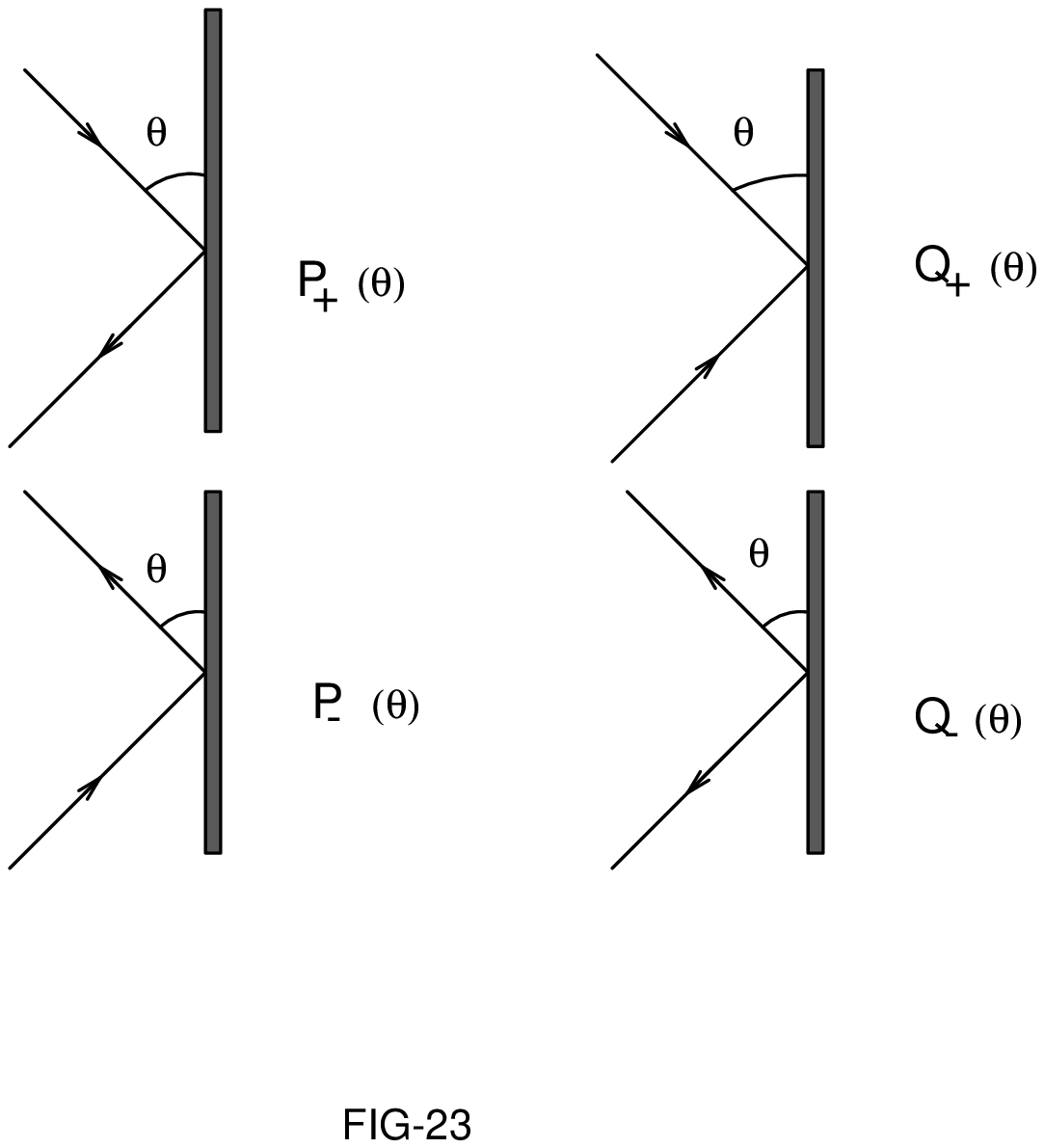,width=4.3in}}}
\vfil\nobreak
\centerline{\hbox{\psfig{figure=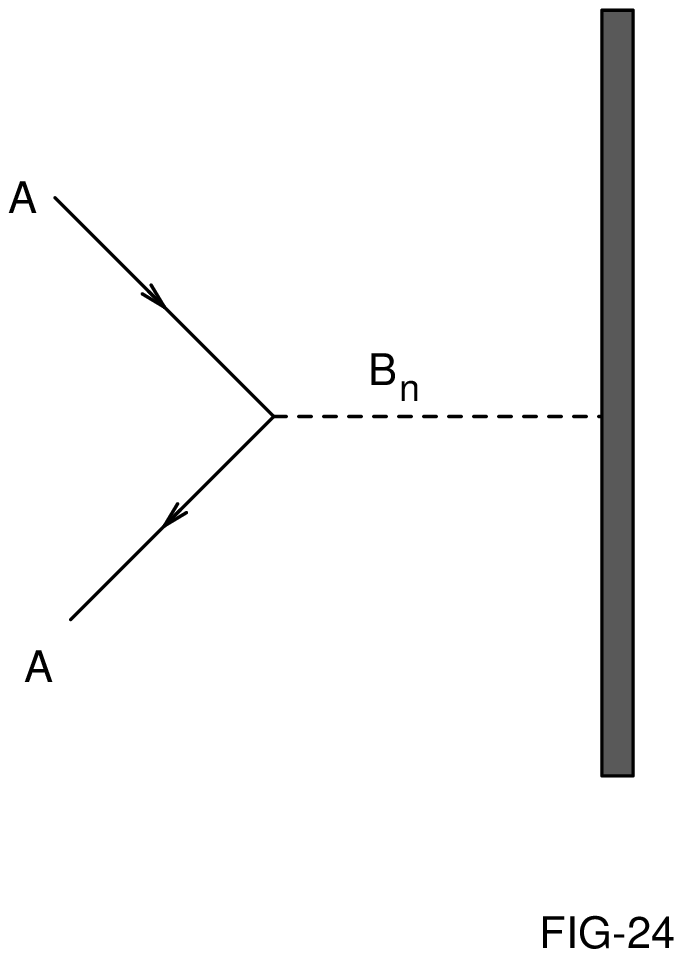,width=2.7in}}}
\break
\centerline{\hbox{\psfig{figure=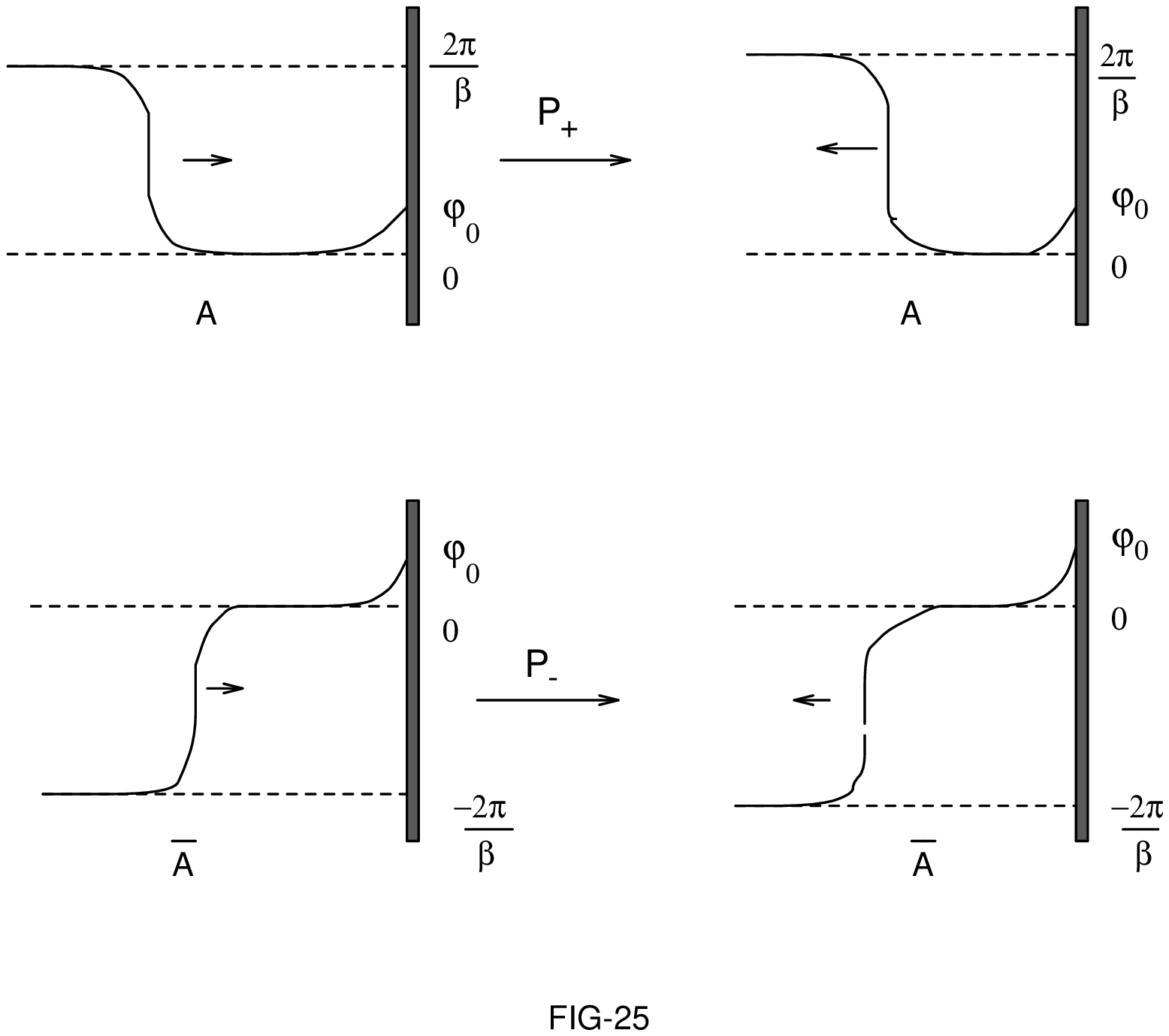,width=7.18in}}}
\end